\documentclass[onecolumn,authoryear]{els-mrw} 

\usepackage{amsmath,amssymb,amsfonts,amsthm,makeidx,graphicx}
\usepackage{txfonts}
\usepackage{helvet}

\usepackage{orcidlink}
\usepackage{xspace}
\usepackage{aas_macros}

\newcommand{\msun}{\mbox{M$_\odot$}}
\newcommand{\yr}{\mbox{${\rm yr}$}}
\newcommand{\myr}{\mbox{${\rm Myr}$}}
\newcommand{\gyr}{\mbox{${\rm Gyr}$}}
\newcommand{\pc}{\mbox{${\rm pc}$}}

\newcommand{\kpc}{\mbox{${\rm kpc}$}}
\newcommand{\kms}{\mbox{${\rm km}~{\rm s}^{-1}$}}

\newcommand{\kcmc}{\mbox{${\rm K}~{\rm cm}^{-3}$}}

\newcommand{\feh}{\mbox{$[{\rm Fe}/{\rm H}]$}}

\newcommand{\be}{\begin{equation}}
\newcommand{\ee}{\end{equation}}
\newcommand{\bea}{\begin{eqnarray}}
\newcommand{\eea}{\end{eqnarray}}

\begin{document}

\chapter{The Formation of Globular Clusters}\label{chap1}

\author[1,2]{J.~M.~Diederik Kruijssen}\orcidlink{0000-0002-8804-0212}%

\address[1]{\orgname{Technical University of Munich}, \orgdiv{Department of Aerospace and Geodesy, Chair of Remote Sensing Technology}, \orgaddress{Arcisstr.~21, 80333 Munich, Germany}}
\address[2]{\orgname{Cosmic Origins Of Life (COOL) Research DAO}, \orgaddress{{https://coolresearch.io}}}

\articletag{Chapter Article tagline: update of previous edition,, reprint..}

\maketitle

\begin{abstract}[Abstract]
  Globular clusters (GCs) are among the oldest and most luminous stellar systems in the Universe, offering unique insights into galaxy formation and evolution. While the physical processes behind their origin have long remained elusive, major theoretical and observational developments in the past decade have led to a new understanding of GCs as the natural outcome of high-pressure star formation in high-redshift galaxies. This review synthesizes recent advancements in our understanding of GC formation and aims to provide a comprehensive point of reference for leveraging the revolutionary capabilities of the current and upcoming generation of telescopes. The latest generation of GC models combine our understanding of their formation and destruction with advanced galaxy formation simulations. The next decade will provide the first-ever opportunity to test such models across their full evolutionary history, from GC formation at high redshift as seen with the James Webb Telescope, to snapshots of GC demographics at intermediate redshifts obtained with 30m-class telescopes, and eventually to the well-characterized GC populations observed at the present day. We identify the major questions that we should expect to address this way.
\end{abstract}

\begin{keywords}
  Globular star clusters, young massive clusters, star formation, galaxy formation, galaxy evolution, high-redshift galaxies, interstellar medium, James Webb Space Telescope
\end{keywords}

\begin{glossary}[Glossary]
  \term{Blue tilt} The observed trend of increasing minimum GC metallicity with increasing GC luminosity in nearby galaxies.\\
  \term{Cluster formation efficiency (CFE)} The fraction of stars that form in gravitationally bound clusters relative to the total stellar mass formed within a galaxy over a certain time period.\\
  \term{Extremely Large Telescope (ELT)} Ground-based observatory with a 39~m primary mirror, currently under construction.\\
  \term{[Fe/H]} Logarithm of the iron-to-hydrogen mass ratio relative to the solar composition.\\
  \term{Fossil stream} A stream of stars in a galaxy's halo that is thought to be the remnant of a disrupted dwarf galaxy or GC.\\
  \term{`Galactic' vs.\ `galactic'} Being part of the Milky Way vs.\ being part of a galaxy in general.\\
  \term{Globular cluster (GC)} It is... complicated. As explained in the text, the author prefers to define this as a gravitationally bound, stellar cluster that in terms of its position and velocity vectors does not coincide with the presently star-forming component of its host galaxy. This is effectively a condition for long-term survival. However, there exists a rich history of definitions for GCs, and the reader is referred to the text for a detailed discussion.\\
  \term{GC mass function (GCMF)} The mass distribution of all GCs in a galaxy.\\
  \term{Giant molecular cloud (GMC)} A cloud of cold molecular gas that resides in the ISM of galaxies.\\
  \term{Hubble Space Telescope (HST)} Optical/near-infrared/near-ultraviolet space-based observatory, launched in 1990.\\
  \term{Hubble time} The age of the Universe.\\
  \term{Initial cluster mass function (ICMF)} The initial mass distribution of star clusters that form within a galaxy over a certain time period.\\
  \term{Interstellar medium (ISM)} The gas and dust that permeates galaxies.\\
  \term{James Webb Space Telescope (JWST)} Infrared space-based observatory, launched in 2021.\\
  \term{Metallicity} The abundance of elements heavier than helium relative to hydrogen.\\
  \term{Metallicity floor} The approximate minimum metallicity (of $\feh\sim-2.5$) that is observed in GC populations.\\
  \term{Multiple populations} The presence of stars with different chemical compositions in a GC.\\
  \term{Proto-GC} A young massive cluster observed at high redshift that might be the precursor of a long-lived GC.\\
  \term{Recombination} The epoch during which electrons and protons first became bound to form neutral hydrogen atoms.\\
  \term{Redshift} The ratio of the observed wavelength to the emitted wavelength of a photon. In cosmology, it reflects the age of the Universe at the time of emission.\\
  \term{Specific frequency} The number of GCs per unit galaxy mass or luminosity.\\
  \term{Star formation efficiency (SFE)} The fraction of the gas mass that is converted into stars.\\
  \term{Stellar feedback} The energy and momentum injected into the interstellar medium by stars.\\
  \term{Tidal shock} The injection of kinetic energy into a star cluster by a transient tidal perturbation, such as a close encounter with a giant molecular cloud.\\
  \term{Young massive cluster (YMC)} A star cluster with a mass in excess of $10^4~\msun$ that formed less than 100~Myr ago.
\end{glossary}

\section*{Learning Objectives}
\begin{enumerate}
\item Obtain a comprehensive overview of the long history of globular cluster formation studies.
\item Gain insight into how and why recent studies have been converging on a new understanding of GCs as the natural outcome of regular star formation in high-redshift galaxies.
\item Learn about the physical conditions required for globular cluster formation and survival.
\item Learn how the demographics of globular clusters emerged and how these trace the physics of their origin.
\item Understand the emergence of globular clusters in the context of galaxy formation and evolution.
\item Collect new ideas for future research by identifying and addressing the main open questions in the field of globular cluster formation.
\end{enumerate}

\section{Introduction: Globular Clusters as Survivors from a Distant Past} \label{sec:intro}

\subsection{The Discovery of Globular Clusters}\label{sec:context}

The first observations of globular clusters (GCs) date back to antiquity, but it took until the 17th century before they were recognized as more than individual stars \citep{halley1679,halley1716,kirch1682}. At the time, they were referred to as \emph{nebul\ae} or bright spots like clouds \citep{halley1716}. Nearly a century passed until \citep{herschel89} realized that these objects could be resolved into collections of thousands of stars. He coined the term \emph{globular cluster} to describe them, and hypothesized that a central, attractive force could explain their spherical shape, causing them to become more compact as they age.

More than two centuries since Herschel's discovery, over 150 GCs have been identified in the Milky Way\footnote{See e.g.\ \citet{harris96,harris10} and \href{https://people.smp.uq.edu.au/HolgerBaumgardt/globular/}{https://people.smp.uq.edu.au/HolgerBaumgardt/globular/}.}, and the arrival of large-scale observational facilities in the 20th century has resulted in the discovery of GC populations in effectively all galaxies with stellar masses above $10^9~\msun$ \citep[e.g.][]{brodie06}. Thanks to extensive theoretical and numerical work, we now know that the mutual gravitational attraction between the stars in a GC is indeed the dominant force shaping their structure and evolution \citep[e.g.][]{heggie03}, and the development of advanced stellar evolutionary models has established that GCs are among the oldest objects in the Universe, with ages spanning a Hubble time \citep[e.g.][]{marinfranch09,forbes10,dotter11,vandenberg13}. The old age of GCs has triggered two active and long-lasting branches of research: the study of the formation and evolution of GCs or GC populations in the context of their host galaxies \citep[e.g.][]{peebles68,fall85,li18,pfeffer18,keller20}, and their use as fossils to reconstruct the formation and assembly histories of their host galaxies \citep[e.g.][]{searle78,kruijssen19d,kruijssen19e}. While this review will mostly focus on the formation and evolution of GCs, we will also briefly touch on how the insights from GC formation research imply that they are highly suitable archaeological tools for galaxy formation studies.

While their origin is an active topic of research, GCs additionally represent fundamental astrophysical laboratories for the study of stellar dynamics \citep[e.g.][]{heggie03,baumgardt03}, stellar evolution \citep[e.g.][]{gratton12,bastian18}, black holes and gravitational waves \citep[e.g.][]{antonini19}, dark matter \citep[e.g.][]{baumgardt09,conroy11b}, and even cosmology and reionization \citep[e.g.][]{ricotti02}. This review focuses on the question of GC formation, and we refer the reader interested in other aspects of GC research to the above references. We also acknowledge earlier reviews that provide important context for this work \citep[e.g.][]{brodie06,kruijssen14c,bastian18,forbes18,krumholz19,adamo20}.

\subsection{What Is a Globular Cluster?} \label{sec:definition}
\begin{figure}[t]
  \centering
  \includegraphics[width=\textwidth]{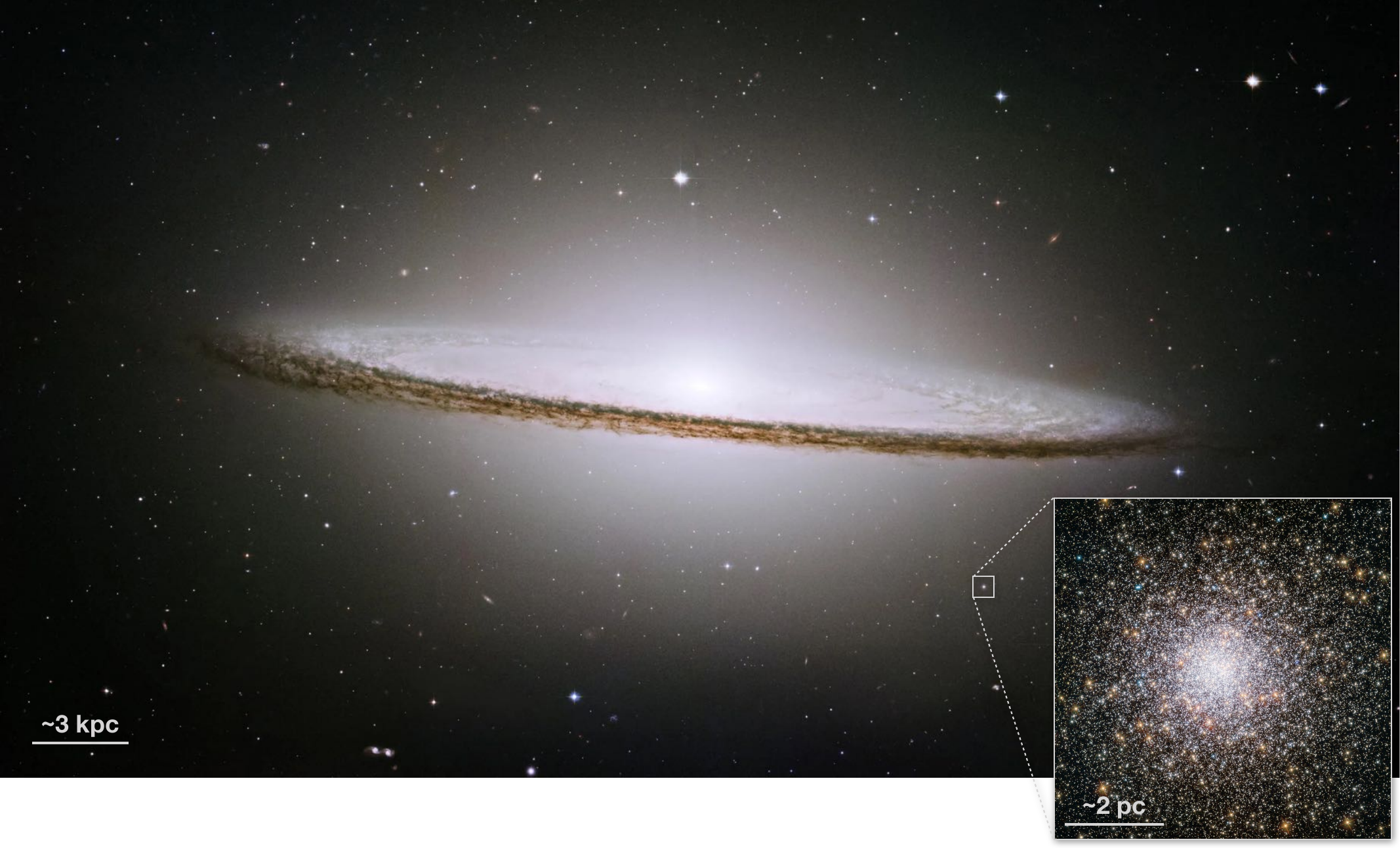}
  \caption{HST image of the Sombrero Galaxy (M104), located at a distance of 9.5~Mpc \citep{mcquinn16}. Many of the bright dots scattered throughout the galaxy are GCs. The inset illustrates this by showing an image of the Galactic GC NGC~362 (not located in M104, but in the Milky Way at a distance of just 8.9~kpc, see \citealt{gontcharov21}). Scale bars indicate the enormous dynamic range in spatial scales spanned by the physical environments relevant to understanding GC formation and evolution. Main image credit NASA and The Hubble Heritage Team (STScI/AURA). Inset credit ESA/Hubble \& NASA.}
  \label{fig:m104}
\end{figure}

No single definition of GCs (illustrated in \autoref{fig:m104}) has been converged on by the research community, and this is largely a result of the historical context of their discovery. The bias imparted by first discovering GCs in the Milky Way (and specifically in the direct vicinity of the Sun) has implied definitional biases that are often challenged by new discoveries. GCs have been defined by setting limits in metallicity (e.g.\ `metal-poor'), mass (e.g.\ $M=10^4{-}10^6~\msun$), age (e.g.\ $\tau>10~\gyr$), location (e.g.\ within a galaxy's bulge or halo), or chemical homogeneity (`multiple populations'). Exceptions to all of these definitions exist, resulting in a you-know-one-when-you-see-one definition of GCs. Retaining all other of these criteria, there exist GCs that are metal-rich \citep[e.g.][]{harris10}, have masses of $10^3{-}10^4~\msun$ \citep[e.g.][]{mandushev91,mclaughlin05}, are younger than $100~\myr$ \citep[e.g.][]{longmore14}, currently reside in the galactic bulge or disk \citep[e.g.][]{minniti17,gran19}, and are chemically homogeneous \citep[e.g.][]{bastian18}. Some systems, such as NGC6791, have historically even been (mis)classified as open clusters \citep[see e.g.][]{baade31,kinman65,salaris04}, despite having an age \citep[8~Gyr,][]{brogaard21}, metallicity \citep[$\feh=0.4$,][]{carraro06}, orbit \citep[out-of-the-plane,][]{jilkova12}, mass \citep[$5000~\msun$,][]{platais11}, and estimated initial mass \citep[$4\times10^5~\msun$][]{dalessandro15} that are entirely consistent with GC populations in the Milky Way and other galaxies \citep[e.g.][]{harris96,dinescu99,forbes10,reinacampos18,pfeffer23}.

Clearly, the definition of GCs is not simple because it is colored by the historical context of their discovery. Galactic GCs are among the brightest objects in the sky, and therefore the first GCs to be discovered were typically bright, massive objects. Because the most intense star formation episodes in the Milky Way's history took place earlier than in most other galaxies of similar mass \citep[e.g.][]{papovich15}, and the Milky Way has been quiescent for most of its history \citep[e.g.][]{kruijssen19e,mackereth19,woody24}, Galactic GCs are mostly $>10~\gyr$ of age. Because they are so old, Galactic GCs are typically more metal-poor than the Sun, or other stars forming today \citep[e.g.][]{snaith15}. Due to the dust extinction in the Galactic plane, most GCs were initially found in the Galactic halo, with more recent surveys finding them in the Galactic bulge \citep[e.g.][]{gran19}. And because the degree of chemical inhomogeneity in GCs increases towards more massive (and potentially older) objects \citep[e.g.][]{carretta10b}, only recently have we started to find GCs that are chemically homogeneous \citep[e.g.][]{martocchia18}.

If we consider the type of object that historically is referred to as a GC, we can understand them as the extreme end of a broad spectrum of stellar clusters that form as a byproduct of the regular star formation process, and survived until the present day. The survival element is key -- the existence of GCs does not just require conditions conducive to the formation of young massive clusters (YMCs), but also for these YMCs to survive and (at some point) become GCs. This perspective led to the definition of \citet{kruijssen15b} that a GC is ``a gravitationally bound, stellar cluster that in terms of its position and velocity vectors does not coincide with the presently star-forming component of its host galaxy.'' The loose quantitative properties of GCs enumerated above (as well as their exceptions) then emerge as a natural part of GC formation and evolution in the context of their host galaxies. In this definition, `stellar cluster' excludes dwarf galaxies, and the `presently star-forming component' is the thin, gas-rich disk where stars are forming today. GCs may traverse the disk, or co-rotate outside of it, but not spend significant time there. \citet{kruijssen15b} reasoned that this definition is useful due to its implicit survival condition: GCs that are durably associated with the gas-rich host galaxy disk are disrupted by strong tidal perturbations from giant molecular clouds (GMCs), so that the migration of GC-like YMCs away from the gas-rich host galaxy disk indeed marks the moment at which they become long-lived GCs. This definition offers a physical perspective and provides concrete predictions, but is somewhat impractical for observational work.

It is an uncomfortable reality that the historical context of discovering GCs has colored our ideas on what GCs are, and also that the Universe does not like quantized categorization the way humans do. Many astrophysical phenomena exist on a continuous spectrum, and GCs are no exception. As we will discuss in this review, GCs likely emerged as part of a continuous star and cluster formation process, throughout the cosmic history of galaxy formation and evolution.

\subsection{Outline of this Review} \label{sec:outline}
This review is organized as follows. After a brief historical overview of GC formation theories in Section~\ref{sec:history}, we will discuss the physics of star cluster formation in Section~\ref{sec:formation}. The destruction and survival of star clusters is reviewed in Section~\ref{sec:destruction}, and the combination of cluster formation and destruction is placed in the context of their host galaxies in Section~\ref{sec:galaxies}. We will conclude with a discussion of current frontiers in GC formation research and some salient concluding perspectives in Section~\ref{sec:frontiers}.

\section{History of Globular Cluster Formation Theories} \label{sec:history}
If we zoom out and consider how the prevalent ideas on GC formation have evolved over the past 60 years or so, we see that these have been quite sensitive to the `flavor of the decade', being inspired by the observational discoveries of the time. The enigmatic nature of GCs has been such a long-standing topic of research, that the introduction of new astrophysical concepts invariably and naturally prompted the idea that maybe this new phenomenon might explain the origin of GCs. Often such GC formation theories relied on special physical conditions, resulting in fine-tuning and incidence problems, but over the past decade a new generation of models emerged that describes GCs as the natural outcome of the regular star formation process under early-Universe conditions.

\subsection{Brief History of `Special Condition' Formation Theories} \label{sec:special}
\textbf{Cosmological conditions in the early Universe.} The 1960s till the late 1980s witnessed rapid advancements in many fields of astrophysics, but few were as impactful as the galaxy formation and cosmology boom, culminating in the cold dark matter (CDM) cosmogeny \citep{white78,davis85}. The strong focus on galaxy formation and the eventual quantitative success of CDM cosmology were accompanied by an increasing awareness of what the conditions in the early Universe were, and how these conditions might have played a role in GC formation. Initial GC formation theories relied on the idea that exotic, early-Universe conditions must have been key to GC formation. \citet{peebles68} hypothesized that GCs could form thanks to the high \citet{jeans02} mass after recombination, whereas \citet{fall85} suggested GCs formed from thermally-unstable, metal-poor gas in hot galactic halos. In the years since, counterevidence accumulated, including the discovery of GCs in dwarf galaxies (which are too low in mass to support hot halos, e.g.\ \citealt{mo10}) such as the Fornax dwarf spheroidal galaxy \citep[e.g.][]{harris81}, and the measurement of GC ages indicating a time of formation several Gyr after recombination \citep[e.g.][]{marinfranch09}.

\textbf{Major galaxy mergers.} Perhaps the biggest revolution that helped overcome the picture of GC formation requiring special, early-Universe conditions was the discovery in the 1980s and 1990s that GC-like YMCs (with masses $10^4{-}10^8~\msun$, radii $0.5{-}10~\pc$, and ages $10^6{-}10^9~\yr$) are still forming today. After early observations with ground-based telescopes \citep{schweizer82,schweizer87}, the launch of the Hubble Space Telescope (HST) provided the sensitivity and resolution needed to discover young GCs throughout rapidly star-forming galaxies in the local Universe \citep{holtzman92,schweizer96,whitmore99,bastian06,adamo20b}. Interestingly, these young GCs were almost exclusively found in galaxy interactions, naturally leading to the idea that GCs formed during major galaxy mergers \citep[e.g.][]{ashman92}. The added advantage of this idea was that the galaxy merger rate peaked in the early Universe \citep[e.g.][]{genel09}, as galaxies assembled through the hierarchical merging of smaller systems \citep{white78}, and therefore the GC formation rate would naturally peak in the early Universe as well. However, the galaxy merger rate also increases with galaxy mass, and the observed increase of the specific frequency (number of GCs per unit galaxy mass or luminosity) towards lower galaxy masses \citep[e.g.][]{peng08} and towards galaxy halo outskirts, where accreted dwarf galaxy debris constitutes a larger fraction of the stellar mass density \citep[e.g.][]{forbes97}, both rule out the idea that most GCs formed during major galaxy mergers.

\textbf{Dark matter halos.} Following the merger hypothesis, new ideas were needed. In the late 1990s and early 2000s, the success of the CDM cosmogeny led to the formulation of `universal' dark matter halo profiles \citep{navarro96,navarro97} and increasingly large-scale simulations of the dark matter distribution in the Universe \citep{springel05d}. On the wings of this success, the idea that GCs might have formed in their own dark matter halos gained traction \citep[e.g.][]{bromm02,bekki06,boley09}. The absence of conclusive evidence for dark matter in the weak gravitional field of GC outskirts \citep[e.g.][]{baumgardt09,conroy11b} prompted the next generation of dark matter halo-based GC formation theories, in which the dark matter is removed by tidal stripping \citep[e.g.][]{trenti15} or head-on collisions between the gaseous GC progenitor structures \citep[e.g.][]{mandelker18,madau20}. Both of these ideas run into order-of-magnitude incidence problems, as the required conditions are much too rare to explain the observed GC population \citep[e.g.][]{keller20}. Additionally, most GCs do not exhibit the metallicity spreads expected in this scenario \citep{bastian18}. In summary, it does not seem feasible that GCs formed in their own dark matter halos.

\textbf{Former nuclear star clusters.} A fourth class of GC formation theories emerged in the late 2000s and early 2010s, due to the discovery of ultra-compact dwarf galaxies (UCDs; e.g.\ \citealt{hilker99,hilker07,drinkwater03}). These form a natural extension of GCs in terms of their structural properties \citep[e.g.][]{misgeld11}, but do contain dark matter \citep[e.g.][]{misgeld11} and are chemically inhomogeneous \citep[e.g.][]{carretta10,bastian18,sills19}. The prevalent view is that these properties are naturally explained if UCDs are former nuclear clusters of tidally stripped dwarf galaxies \citep[e.g.][]{pfeffer13}, leading to the idea that at least some GCs might be former nuclear clusters too \citep[e.g.][]{mackey05,lee09}. However, the number of UCDs is considerably too low to explain the observed GC population \citep[e.g.][]{pfeffer14,pfeffer21,kruijssen19e} and most GCs lack the heavy element abundance spreads observed in former nuclear clusters \citep[e.g.][]{bastian18}.

\subsection{Current Perspective: Globular Clusters as the Relics of Regular Star Formation in High-Redshift Galaxies} \label{sec:regular}
In the early 2010s, the increasing capabilities of sub-mm telescopes led to the first characterizations of the interstellar medium (ISM) in galaxies at the redshifts ($z=2{-}3$) at which GCs should have formed according to their measured ages. These direct detections of molecular gas in the ISM of star-forming galaxies at $z\gtrsim2$ revealed that the high ISM pressures found in local-Universe galaxy mergers with widespread YMC formation \citep[$P/k\gtrsim10^7~\kcmc$, e.g.][]{wei12} are prevalent in the normal star-forming disks of galaxies at the epoch of GC formation \citep[e.g.][]{genzel10,shapiro10,tacconi13,tacconi18,swinbank11,swinbank12}. In other words, galaxies near the peak of the cosmic star formation rate density at $z\sim2{-}3$ \citep[e.g.][]{madau14} were found to host exactly those conditions that are conducive to the formation of YMCs (and, if they survive, potential GCs).

As a result of these discoveries, the obvious questions to ask are: could the products of regular cluster formation at high redshift have survived until the present day? And if so, are these relics consistent with the properties of local GC populations? These are simple questions to ask, but they are hard to answer because they require an end-to-end model for GC formation and evolution in the context of galaxy formation and evolution. More concretely, they require a model for star formation, for the associated initial demographics of the cluster population, for the physical processes that drive cluster evolution and destruction, for the environmental dependence of these processes, and for the (re)distribution of clusters during galaxy formation and assembly.

Pioneering studies in the 1990s and 2000s had already suggested that GCs should naturally form from high-pressure molecular clouds in the ISM of galaxies, irrespective of whether these galaxies were merging or not \citep{harris94,elmegreen97,kravtsov05}. The systematic characterization of the ISM in high-redshift galaxies subsequently triggered an explosion of theoretical work focused on modelling the origin of GCs in the context of star formation in their host galaxies \citep[e.g.][]{kruijssen15b,li17,li18,li22,choksi18,pfeffer18,keller20,reinacampos22,grudic23,rodriguez23,delucia24}, with considerable success in reproducing the observed demographics of GC populations.

\begin{figure}[t]
  \centering
  \includegraphics[width=\textwidth]{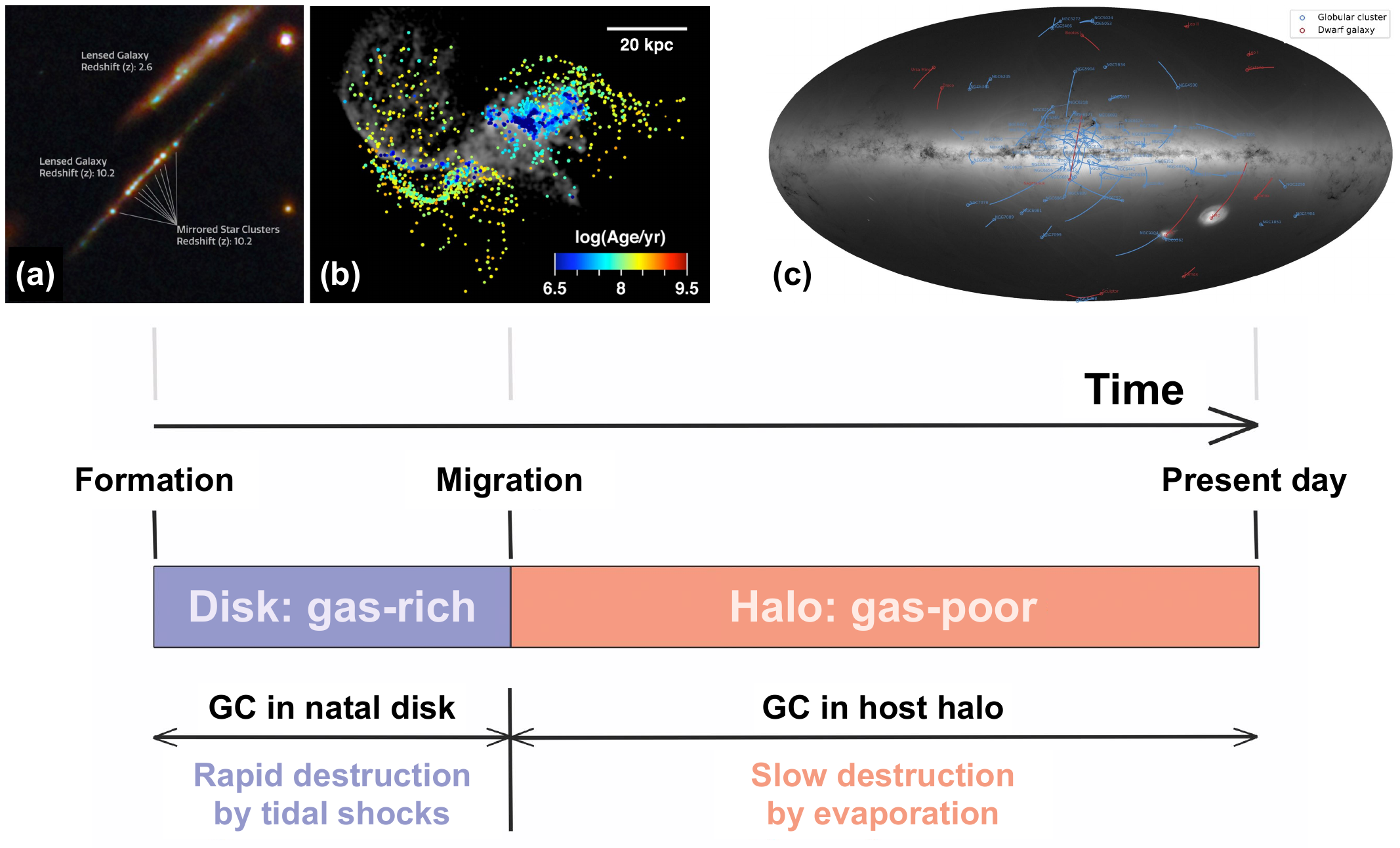}
  \caption{Schematic representation of the ingredients contained in modern GC formation models that assume GCs are the natural outcome of regular star formation under high-pressure conditions (adapted from \citealt{kruijssen15b}). Initially, stars and stellar clusters form through the regular star formation process in the gas-rich disks of (high-redshift) galaxies (see panel a). During their subsequent evolution within the natal galaxy disk, the clusters undergo a phase of rapid destruction by tidal perturbations from dense gas concentrations in the disk. This phase leads to the destruction of most of the low-mass clusters and lasts until the remaining, generally more massive clusters migrate out of this disruptive environment, plausibly by galaxy mergers or tidal stripping (see panel b). This migration considerably increases their long-term probability of survival, and marks the moment where they become long-lived GCs. These GCs undergo further quiescent dynamical evolution until the present day, during which they exclusively lose stars by the process of gradual dynamical evaporation within the host galaxy's potential. GCs in the Milky Way are currently in this phase of their evolution (see panel c). Panel (a) shows the gravitationally bound YMCs in the lensed Cosmic Gems arc, with YMCs concentrated within regions estimated to be less than 50~pc across \citep[image credit ESA/Webb, NASA \& CSA, L.~Bradley, A.~Adamo and the Cosmic Spring collaboration]{adamo24}. Panel (b) shows the migration of stellar clusters in a galaxy merger model from \citet{kruijssen12c}, with colors indicating the cluster ages. Panel (c) shows the present-day GC population of the Milky Way, with blue circles and lines visualizing GCs and their orbits \citep[image credit ESA/Gaia/DPAC, A.~Helmi, M.~Breddels, F.~van Leeuwen, P.~McMillan]{gaia18b}.}
  \label{fig:architecture}
\end{figure}
The latest generation of GC formation models collectively approximately follow the architecture shown in Figure~\ref{fig:architecture} \citep{kruijssen15b}. Initially, stars and stellar clusters form through the regular star formation process in the gas-rich disks of (high-redshift) galaxies \citep[e.g.][]{vanzella17,vanzella22,adamo24}. The star-forming environments of these galaxies represent the most common form of high-pressure conditions in the history of the Universe and enable an elevated fraction of stars to form in clusters \citep[e.g.][]{elmegreen97,elmegreen08,kruijssen12d}, some of which reach masses $>10^6~\msun$ \citep[e.g.][]{reinacampos17}. During their subsequent evolution within the natal galaxy disk, the clusters undergo a phase of rapid destruction by tidal perturbations from dense gas concentrations in the disk \citep[e.g.][]{gieles06,meng22,webb24}. This phase leads to the destruction of most of the low-mass clusters and lasts until the remaining, generally more massive clusters migrate out of this disruptive environment, plausibly by galaxy mergers or tidal stripping \citep[e.g.][]{kruijssen12c}. This migration considerably increases their long-term probability of survival \citep[e.g.][]{pfeffer18,li22,newton24}, and marks the moment where they become long-lived GCs. These GCs undergo further quiescent dynamical evolution until the present day, during which they exclusively lose stars by the process of gradual dynamical evaporation within the host galaxy's potential \citep[e.g.][]{baumgardt03,gieles08}.

Summarizing, the several elements of Figure~\ref{fig:architecture} each explain different aspects of the origin of GCs. The high-pressure environment shown in panel (a) explains why massive clusters could form in the first place. The phase of rapid destruction by tidal shocks explains why mostly massive GCs are left. The migration of GCs shown in panel (b) explains why most GCs are not associated with the host galaxy disk. Finally, the phase of slow destruction by evaporation, which characterizes the GCs in the Milky Way shown in panel (c), explains why any GCs survived at all.

\subsection{Towards Comprehensive Models for Globular Cluster Formation during Galaxy Formation and Evolution}
Individual parts of the timeline of GC formation and evolution shown in Figure~\ref{fig:architecture} have been studied in great detail, using a variety of modelling techniques. For instance, high-resolution hydrodynamical simulations of high-redshift galaxy formation and galaxy mergers have been able to directly resolve the formation of extremely massive YMCs in the high-pressure ISM \citep[panel a; e.g.][]{li17,kim18,lahen20,ma20}. Semi-analytic models of galaxy formation and evolution have `painted on' GCs to trace their migration and gradual evaporation during galaxy assembly, often with the goal of studying correlations between GC system demographics and galaxy properties \citep[panels b and c; e.g.][]{beasley02,muratov10,tonini13,li14,boylankolchin17,boylankolchin18,choksi18,choksi19}. Similarly, hydrodynamical simulations of galaxy formation have been post-processed through `particle-tagging' techniques, identifying subsets of (stellar or dark-matter) particles that are designated as GCs, enabling insight into the environments in which GCs might have formed, and into how they might have been (re)distributed during the galaxy assembly process \citep[e.g.][]{bekki02,li04,kravtsov05,moore06,mistani16,renaud17,elbadry19,halbesma20,valenzuela21,doppel23}. At first sight, models that tag GCs to particles in dark matter-only simulations may seem to mostly serve a phenomenological purpose, given that GCs must have formed from gas and there exists no fundamental connection to a galaxy's dark matter content. However, they have been quite successful in showing how observed scaling relations between the GC population and galaxy properties can be reproduced with minimal physics \citep[e.g.][]{boylankolchin17,elbadry19,valenzuela21}, thus helping identify which other observables are physically more meaningful. Finally, there exists extensive literature on direct $N$-body simulations of the evolution and destruction of individual star clusters under the influence of time-variable tidal fields. Some of these have focused on mild tidal perturbations due to cluster migration during the galaxy assembly process \citep[e.g.][]{rieder13,renaud13,carlberg18,webb19b}, whereas others have aimed to study rapid cluster destruction by tidal perturbations from dense gas concentrations in the natal galaxy disk \citep[e.g.][]{webb19,webb24,martinezmedina22,lahen24}.

In order to reproduce observed GC populations, the modern generation of GC formation models aims to include all ingredients shown in Figure~\ref{fig:architecture} \citep[e.g.][]{kruijssen15b,pfeffer18,li19,li22,reinacampos22,delucia24}. The first attempt at such comprehensive models was the E-MOSAICS suite of simulations \citep{pfeffer18,kruijssen19d,bastian20}, which combined a physical model for star cluster formation and destruction \citep{kruijssen11,pfeffer18} with the galaxy formation model of the EAGLE simulations \citep{schaye15,crain15}. By including physical descriptions for all important steps and identifying key observables for testing these, such models are highly predictive. While they certainly fail in some aspects, they have helped the field make important progress in improving the physical understanding of GC formation. Among this family of models' key successes is the ability to simultaneously reproduce:
\begin{enumerate}
  \item the observed demographics of young cluster populations in the nearby Universe, e.g.:
  \begin{enumerate}
    \item the cluster formation efficiency \citep[e.g.][]{li17,pfeffer19b,lahen20,li22,grudic23};
    \item the initial cluster mass function \citep[e.g.][]{li17,li18,pfeffer19b,li22,lahen24};
    \item cluster age distributions \citep[e.g.][]{miholics17,pfeffer19b}.
  \end{enumerate}
  \item the observed properties of present-day GC populations after nearly a Hubble time of evolution, e.g.:
  \begin{enumerate}
    \item the specific frequency, i.e.\ the number of GCs per unit galaxy mass or luminosity \citep[e.g.][]{choksi19,kruijssen19d,bastian20};
    \item radial density profiles and kinematics of GC populations \citep[e.g.][]{trujillogomez21,chen22,reinacampos22b,rodriguez23};
    \item GC metallicity distributions \citep[e.g.][]{li14,li19,choksi18,pfeffer23};
    \item GC age and age-metallicity distributions \citep[e.g.][]{li14,li19,kruijssen19d,kruijssen19e,delucia24};
    \item the (high-mass end of the) GC mass function \citep[e.g.][]{pfeffer18,hughes22,reinacampos22};
    \item the association of GCs with fossil streams \citep[e.g.][]{hughes19};
    \item the contribution of GC destruction to the stellar halo \citep[e.g.][]{hughes20,reinacampos20}.
  \end{enumerate}
\end{enumerate}
Additionally, these models provide predictions for the demographics of YMC populations (and potential GC progenitors, or `proto-GCs') in high-redshift galaxies, such as in what environments proto-GCs formed \citep[e.g.][]{keller20}, what the proto-GC formation history is \citep[e.g.][]{reinacampos19,joschko24}, and the luminosity function of proto-GCs \citep[e.g.][]{pfeffer19,pfeffer24}. Alternative GC formation theories that rely on special physical conditions have not yet been able to achieve a similar predictive power, and additionally would need to explain what happened to the many YMCs that must have formed as part of the regular star formation process, and why these would never have turned into GCs.

Despite their successes, modern GC formation models also exhibit clear weaknesses. They struggle to reproduce the observed degree of GC destruction over a Hubble time if they do not include the physics needed to model the cold ISM \citep[e.g.][]{pfeffer18,kruijssen19d}. The graininess of the gravitational potential generated by the cold and substructured ISM has been suggested to be crucial for the rapid tidal destruction of low-mass clusters on theoretical grounds \citep[e.g.][]{elmegreen10,kruijssen15b}. The latest generation of models that does include a cold ISM component now reproduces the dearth of low-mass GCs indicative of their rapid destruction \citep[e.g.][]{reinacampos22}. However, these models are unable to reproduce the star and cluster formation properties of low-redshift galaxies due to the complex nature of star formation and stellar feedback physics in simulations that do resolve the cold ISM. Finally, modelling GC formation and evolution on top of galaxy formation simulations is inherently an extremely multi-scale problem, and the current generation of models generally needs to rely on sub-grid recipes that carry free parameters that are sometimes difficult to constrain and do impact the results \citep[see discussions in e.g.][]{pfeffer18,li19,reinacampos22}.

Owing to its empirical success, the standard paradigm for GC formation is currently that GCs are the natural outcome of regular star formation under the high-pressure conditions ubiquitous in high-redshift galaxies. These conditions also occur in the local Universe, but are restricted to special environments such as galaxy mergers, galaxy centers, and starburst dwarf galaxies, where indeed most of the present-day YMC formation is observed \citep[e.g.][]{kruijssen14c,adamo20b}. In the remainder of this review, we will discuss the physics driving the formation and destruction of GCs in this context, and provide an outlook towards further tests of this perspective on GC formation.

\section{Star Cluster Formation} \label{sec:formation}
In this section, we provide a brief overview of the physical processes that define the initial conditions of the GC population. We discuss the conditions that are required for the star formation process to (locally) produce gravitationally bound stellar populations, how these conditions across a galaxy result in the observed cluster formation efficiency (CFE; the fraction of star formation occurring in bound clusters), and what the initial cluster mass function (ICMF; the initial mass distribution of the stellar clusters that form in a galaxy) looks like.

\subsection{Conditions for Achieving Gravitationally Bound Stellar Populations} \label{sec:boundedness}
Gravitational boundedness is a necessary condition for the formation of GCs -- without it, stellar groups would undergo ballistic dispersal within a dynamical time, which is of the order 10~Myr for typical star-forming GMCs \citep[e.g.][]{sun22,chevance23}. The formation of gravitationally bound stellar clusters is a complex process that is influenced by a variety of physical mechanisms. These include the balance between the gravitational binding energy of the densest parts of hierarchically fragmenting GMCs and their nascent stellar populations, the influence of stellar feedback during the cluster formation process, and the destructive effects of the tidal field of the host galaxy and its substructure.

Empirically, most stars in the Universe do not currently reside in stellar clusters \citep[e.g.][]{lada03,krumholz19}. Attempts to explain this observation quickly identified the importance of stellar feedback, i.e.\ the deposition of energy and momentum by young stars into their natal GMC. The leading idea was that all stars formed in clusters, but that the energy input from massive stars would be sufficient to unbind the majority of the gas in the GMC, thereby halting star formation before any group of stars might become bound without the gravitational potential of the gas \citep[e.g.][]{tutukov78,hills80,lada84,goodwin06}. These early numerical experiments showed that a star formation efficiency (SFE; the fraction of the gas mass that is converted into stars) of $\gtrsim30\%$ is needed to achieve gravitational boundedness \citep[unless the gas is removed very slowly, see e.g.][]{baumgardt07}. These idealized numerical experiments modelled centrally-concentrated clusters with a static background potential that is gradually removed, and omitted the dynamic nature of the turbulent ISM and the spatial variation of the gas removal by stellar feedback.

The idea that the SFE regulates gravitational boundedness is reasonable, but runs into an important problem when applied in the context offered by early numerical experiments. If a high SFE of $\gtrsim30\%$ is needed to achieve gravitational boundedness under impulsive gas removal, then the existence of bound stellar clusters would naively require GMCs to either convert such a considerable fraction of their gas mass into stars, or to remove the gas very slowly (i.e.\ over $\sim10$ dynamical times). Observationally the SFE per free-fall time is of the order a few percent, i.e.\ a few percent of the gas mass is converted into stars on a free-fall time of $\sim10^7$~yr \citep[e.g.][]{krumholz07,leroy17,utomo18,sun23}. In order to achieve a SFE of $\gtrsim30\%$ at such a slow rate, or eject the gas over $\sim10$ dynamical times, GMCs would need to live for $10{-}30$ free-fall or dynamical times, which would imply a GMC lifetime of $\gtrsim100$~Myr. Observed GMC lifetimes are considerably shorter, of the order of a few dynamical times or $10{-}30$~Myr \citep[e.g.][]{kruijssen19b,chevance20,chevance20b,kim22}, and GMCs are observed to disperse within just a few Myr after the emergence of unembedded massive stars \citep[e.g.][]{chevance22}, causing them to achieve integrated SFEs of just a few percent.

The solution is that the SFE must vary as a function of position and spatial scale across a GMC. That way, GMCs can achieve low global SFEs while still forming bound clusters in their densest parts. This insight originally came from early studies of hydrodynamical simulations of star formation in turbulent molecular clouds \citep[e.g.][]{offner09,girichidis12,kruijssen12}, which achieved gas-poor stellar clusters even without stellar feedback. These simulations showed that newborn stars move subvirially with respect to the parent GMC (i.e.\ their kinetic energy is low relative to their mutual binding energy) as they condense out of the highest-density and slowest-moving parts of the GMC. Due to the short free-fall times of these density peaks ($<0.1$~Myr) and further cluster shrinkage due to continued gas accretion onto the newborn stars, they achieve high local SFEs through local gas exhaustion (rather than expulsion). Even though these simulations were still lacking a model for stellar feedback, they already showed that the spatial variation of the SFE across a GMC is a key ingredient for the formation of bound clusters.

\begin{figure}[t]
  \centering
  \includegraphics[width=\textwidth]{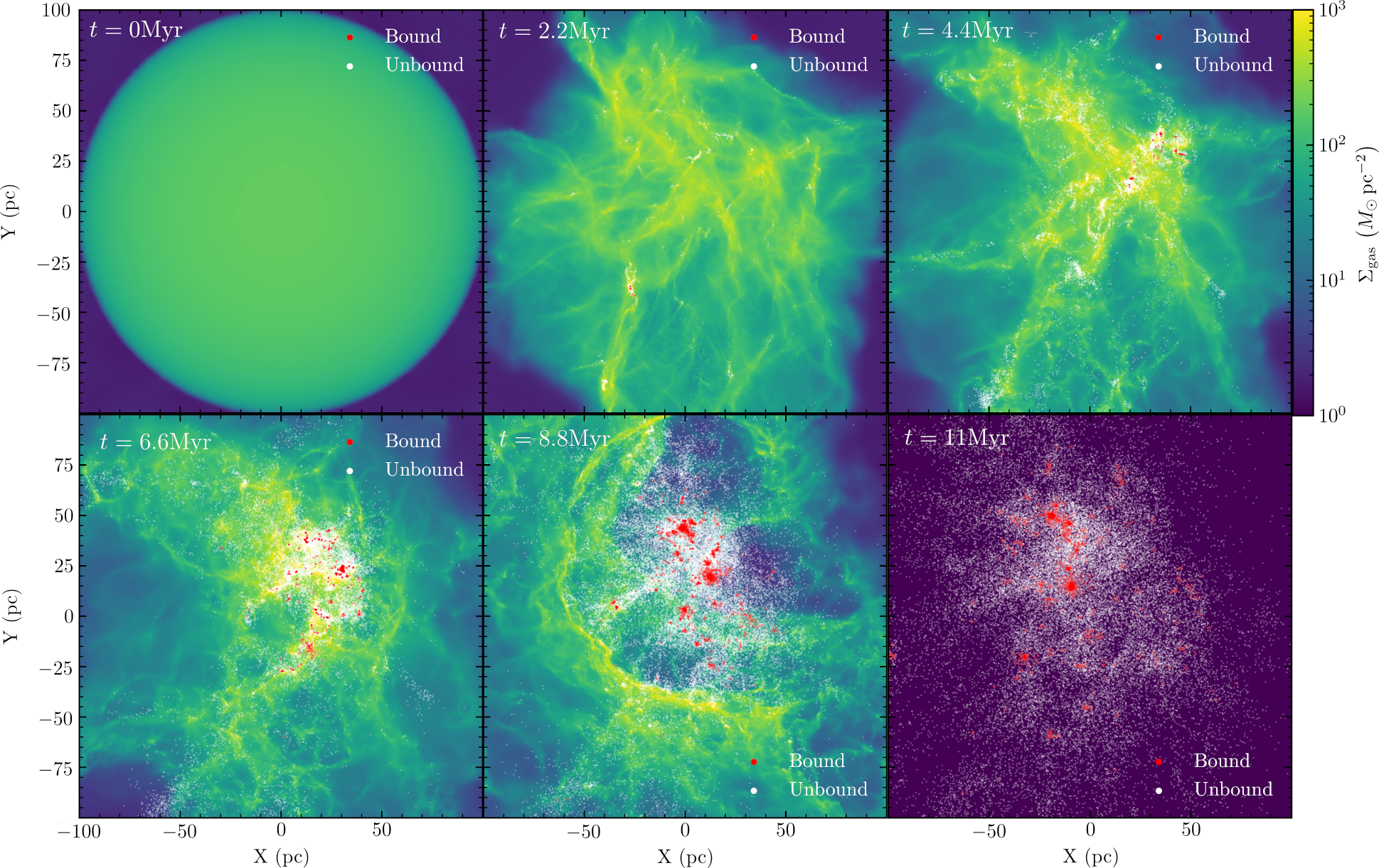}
  \caption{Numerical simulation of star and cluster formation within a GMC of initial mass $4\times10^6~\msun$ and initial size 100~pc. The panels show a sequence of snapshots, with colors indicating the gas surface density (see the color bar) and dots indicating the positions of the newborn stars, with gravitationally bound clusters highlighted in red and the surrounding unbound association shown in white. The times corresponding to each panel are indicated in the upper left corner. The sequence illustrates the hierarchical assembly of bound clusters from the densest parts within a GMC, and how the subsequent feedback-driven gas removal disrupts the global coalescence without unbinding the bound clusters. The total bound fraction achieved this way is around 10\%. Taken from \citet{grudic21}, reproduced with permission.}
  \label{fig:grudic21}
\end{figure}

Modern numerical simulations with comprehensive descriptions of the ISM, self-gravity, star formation, and a variety of stellar feedback mechanisms now provide a more detailed view of the conditions leading to bound cluster formation \citep[illustrated in \autoref{fig:grudic21}; also see e.g.][]{dale15,howard18,li19b,grudic21}. These simulations confirm the idea that the SFE varies strongly with position within the GMC, and that the feedback-driven cessation of star formation determines how far advanced the hierarchical assembly of young stellar groups into gravitationally bound structures \citep[e.g.][]{allison09,maschberger10,fujii12} can proceed before the global gravitational coherence of the natal GMC is disrupted by feedback. As such, stellar feedback does not control the local fraction of bound star formation (which can range from less than a percent to close to 100\% in the most extreme overdensities), as originally postulated by the idealized numerical experiments predating early 2010s, but rather determines the extent to which these overdensities might have coalesced into bound clusters \citep[referred to as `conveyor-belt' cluster formation by][]{longmore14}.

The above physical picture is supported by extensive observational evidence. YMCs themselves are gravitationally bound objects, and lack the kinematic expansion signatures that would be expected if the local SFE would have been low \citep[e.g.][]{henaultbrunet12}. Additionally, the typical sizes of YMCs are much smaller than the sizes of GMC overdensities of similar mass, indicating bound clusters form from contraction rather than expansion \citep[e.g.][]{walker15,walker16}. Observationally inferred GMC lifetimes are too short for them to reach the high SFEs needed for bound cluster formation anywhere but in their densest fragments \citep[e.g.][]{kruijssen19b,chevance20,kim22}. And finally, stellar feedback is observed to be ineffective at halting star formation in the densest parts of GMCs hosting massive star formation \citep[e.g.][]{ginsburg16}, indicating that these proto-cluster environments will achieve gas exhaustion and form bound clusters. A key implication of these results is that not all stars form in bound clusters. Therefore, it is of crucial importance to quantify what fraction of star formation does.

\subsection{The Fraction of Star Formation Occurring in Bound Stellar Clusters} \label{sec:cfe}
With an understanding of the conditions needed for bound cluster formation, it is possible to define the cluster formation efficiency (CFE) as the fraction of star formation in a region that occurs in the form of bound clusters \citep{bastian08}. The CFE can be defined across an entire galaxy (or some fraction thereof containing a significant number of GMCs), and both observationally and theoretically it can be related to the properties of the ISM, which define the initial conditions of the star formation process.

Observationally, it can be challenging to measure the gravitational boundedness of young stellar populations, and infer the CFE robustly. This complication has dominated much of the discussion in young cluster population studies \citep[see the review by][]{krumholz19}, but there are ways in which a bound cluster sample can be defined with some confidence. First, when kinematic information is available, it is possible to select bound systems by requiring that they are older than their dynamical time \citep{gieles11}. Unbound stellar associations expand ballistically, causing their dynamical time to increase at the same rate as their age. However, for extragalactic cluster populations, the kinematic information needed to define the dynamical time is often unavailable. A second way of selecting bound clusters in extragalactic environments is to select centrally concentrated objects with ages of $10{-}50$~Myr \citep{kruijssen16}, after which unbound associations should no longer be confused with the concentrated morphologies of bound clusters \citep{kuhn19,ward20,wright23}. Given that the CFE is defined shortly after cluster formation, cluster samples that extend to older ages are affected by cluster destruction, which biases the CFE towards lower values especially in more disruptive, high-pressure environments (see Section~\ref{sec:destruction}).

When these selection criteria are applied, the CFE is found to be a function of the gas or star-forming properties of the host galaxy. Early studies already identified a trend of the CFE increasing with the star formation rate (SFR) surface density \citep[e.g.][]{goddard10,adamo10,adamo11,cook12}, and later work conclusively showed that this relationship is statistically robust over a large sample of young cluster populations in the local Universe \citep[e.g.][]{adamo15b,adamo20b,adamo20,johnson16,ginsburg18,messa18} and also seems consistent with the first observations of cluster formation at high redshifts \citep[e.g.][]{vanzella22}. It is clear that restricting the cluster sample to compact structures with young (but not too young) ages probes interesting physical processes. But why does a larger fraction of star formation occur in bound clusters in galaxies with higher SFR surface densities?

\begin{figure}[t]
  \centering
  \includegraphics[width=\textwidth]{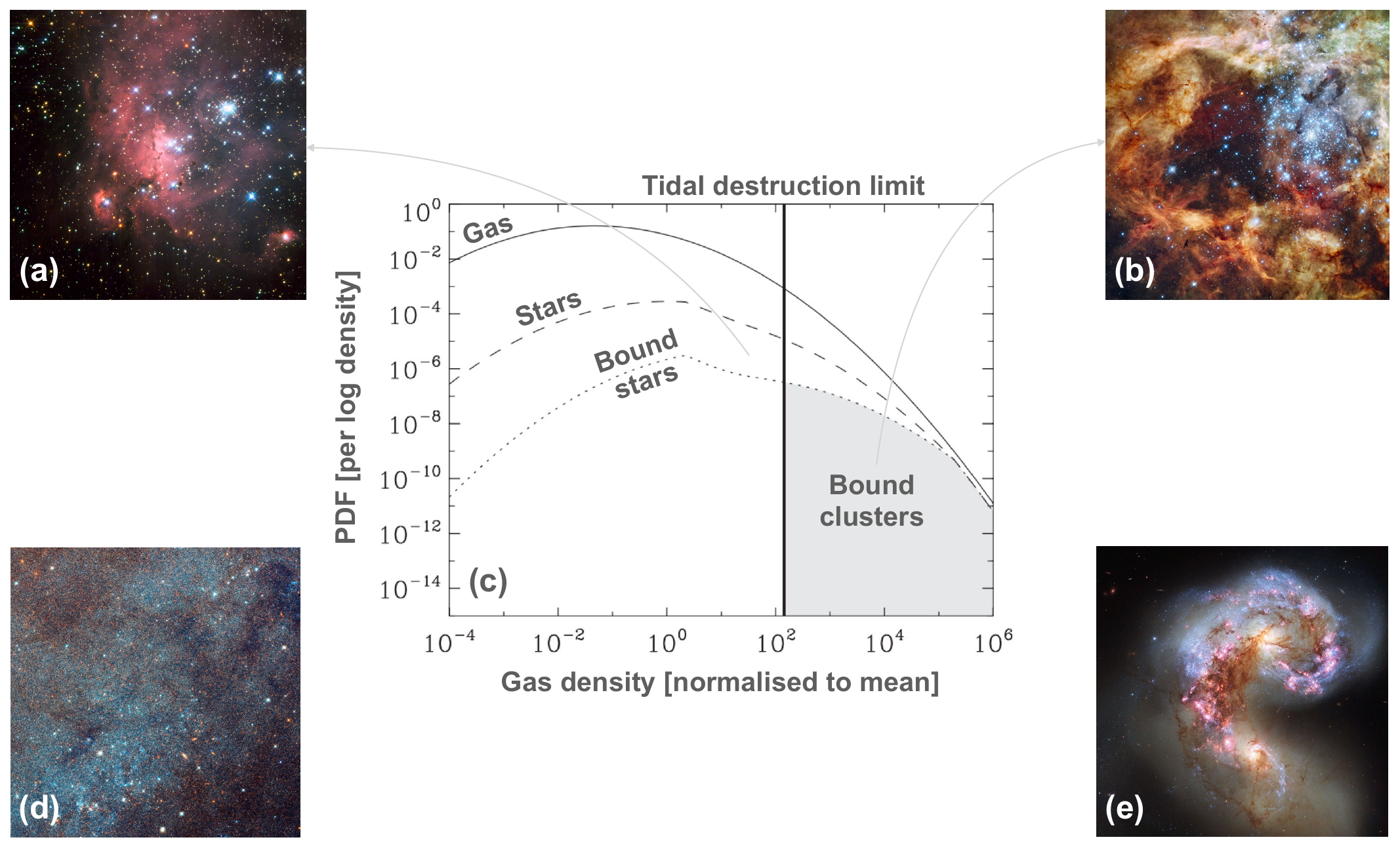}
  \caption{Illustration of the relation between the ISM density PDF and the CFE. Panel (a) shows the OB association LH72 in the Large Magellanic Cloud (image credit ESA/Hubble, NASA and D.~A.~Gouliermis). Panel (b) shows the YMC R136, also in the Large Magellanic Cloud (image credit NASA, ESA, F.~Paresce, R.~O'Connell, and the Wide Field Camera 3 Science Oversight Committee). Panel (c) shows where the formation of these types of systems occurs in the context of the ISM density PDF \citep[taken from][]{kruijssen12d}. At higher densities, the gravitational free-fall time is shorter, leading to a higher integrated SFE, and a higher gravitationally bound fraction of young stars. The vertical line indicates the minimum density contrast needed for surviving tidal perturbations from nearby density structure. The ratio between the integral over the grey-shaded area and the area under the dashed curve yields the CFE. Because the PDF widens under high-pressure conditions, the CFE is higher in such environments too, facilitating potential GC formation. Panel (d) illustrates the low-pressure environment of M31, where a narrow density PDF leads to a low CFE (image credit NASA, ESA, J.~Dalcanton, B.~F.~Williams, L.~C.~Johnson and the PHAT team). Panel (e) shows the high-pressure environment of the Antennae Galaxies, where a broad density PDF leads to a high CFE (image credit ESA/Hubble \& NASA).}
  \label{fig:cfe}
\end{figure}
\citet{elmegreen08} proposed that the CFE can be related to the gas volume density spectrum of the ISM, with the high-density end resulting in bound clusters. This idea was formalized in the cluster formation theory of \citet{kruijssen12d}, wherein the bound fraction of young stellar populations increases towards higher gas volume densities. In the \citet{kruijssen12d} cluster formation theory, the CFE can be obtained by integrating bound cluster formation over the probability distribution function (PDF) of the gas volume density in the ISM after applying simple descriptions of density-dependent star and cluster formation (see \autoref{fig:cfe}). This PDF initially follows a log-normal shape, of which the median increases with the mean gas density and the width increases with the Mach number of the gas \citep[e.g.][]{vazquez94,padoan97,federrath10}. Towards higher gas volume densities, the gas free-fall time decreases. If we adopt an approximately constant SFE per free-fall time \citep[e.g.][]{krumholz05,padoan11}, and assume that gravitational collapse and star formation are halted when the pressure generated by stellar feedback balances the ambient gas pressure, it is possible to derive the SFE, which will increase towards higher gas volume densities. Because the fraction of stars that form in gravitationally bound substructures increases with the SFE \citep[e.g.][]{kruijssen12,li19b,grudic21}, this also means that the prevalence of bound groups will increase towards higher gas volume densities.

Given that the gas density PDF widens and shifts to higher densities for higher mean volume densities and Mach numbers (which together determine the gas pressure as $P\propto\rho{\cal M}^2$), the CFE (obtained as the integral of the SFE over the gas density PDF) must increase towards higher gas pressures. The gas pressure is well-known to be correlated with the gas surface density and the SFR surface density in galaxy disks \citep[e.g.][]{blitz06,ostriker10,krumholz12}, and indeed this correlation can be explained naturally in the context of hydrostatic equilibrium disk modelling \citep[e.g.][]{krumholz05}. Therefore, the natural prediction is that the CFE increases with the gas and SFR surface densities, in accordance with the observations.

Quantitatively, the observed and predicted CFEs range from $\lesssim10\%$ in the low-pressure environments within local-Universe galaxy disks, with gas surface densities below $\sim10~\msun~\pc^{-2}$, to $\sim50\%$ in the high-pressure environments of nearby galaxy centres, galaxy mergers, and high-redshift galaxy disks, with gas surface densities above $\sim300~\msun~\pc^{-2}$. This means that most GCs formed at a time that galaxies were extraordinarily efficient at forming gravitationally bound clusters. The elevated CFE in high-pressure environments also contributes to explaining why GC-like clusters (YMCs) are seen to form in galaxy mergers and starbursts in the local Universe.

\subsection{The Initial Cluster Mass Function} \label{sec:icmf}
Having established the fraction of star formation occurring in bound clusters, it is necessary to define the initial cluster mass function (ICMF; the initial mass distribution of the stellar clusters that form in a galaxy). The properties of the ICMF are affected by the physical conditions in the ISM in a way similar to the CFE. Because the rate of cluster destruction depends on the cluster mass (see Section~\ref{sec:destruction}), understanding how the ICMF varies with environment is key to understanding how the present-day GC population might have originated.

Empirically, the ICMF follows a power law with an index of $\alpha=-2$ (adopting the definition ${\rm d}N/{\rm d}M\propto M^{\alpha}$) over most of the mass range where young clusters are observed \citep[e.g.][]{zhang99,hunter03,larsen09,krumholz19}. This power law shape emerges naturally during scale-free hierarchical gravitational collapse \citep[e.g.][]{elmegreen96} whenever the different scales are initially uncorrelated \citep{guszejnov18}. As a result, a power law with slope $\alpha=-2$ describes self-gravitating structures in the Universe over nearly 20 orders of magnitude in mass, across stars, protostellar cores, GMCs, star clusters, and the dark matter halos of galaxies. Star clusters are no exception, and their mass function only deviates from a power law shape after considerable evolution, because the rate of cluster destruction is correlated with the cluster mass.

The $\alpha=-2$ slope of the ICMF implies that the total mass formed as part of the ICMF is proportional to $\ln{(M_{\rm max}/M_{\rm min})}$, where $M_{\rm max}$ and $M_{\rm min}$ are the maximum and minimum cluster masses in the ICMF. This means that the total mass is infinite if the ICMF extends to arbitrarily low or high masses, and that the cluster population contains equal mass across equal logarithmic mass intervals. To match observations, the power-law ICMF was initially extended with a \citet{schechter76} type exponential truncation at the high-mass end \citep[e.g.][]{gieles09}. A more comprehensive form of the ICMF must adopt (exponential) truncations on both ends, which better describes the observed mass function of young clusters \citep[hereafter `double-Schechter']{trujillogomez19}. The double-Schechter form of the ICMF is given by
\be
\label{eq:icmf}
\frac{{\rm d}N}{{\rm d}M} = M^{\alpha}\exp{\left(-\frac{M_{\rm min}}{M}-\frac{M}{M_{\rm max}}\right)} ,
\ee
where the exponential term ensures that the ICMF extends to finite values at both low and high masses.

In the context of GC formation, it is particularly important to understand the physics that determine the upper end of the ICMF. After all, GCs tend to have relatively high masses compared to clusters forming today, as high-mass clusters are more likely to survive over a Hubble time than low-mass clusters. In environments where the high-mass truncation is too low, no long-lived GCs can form. In the local Universe, observations show that the upper mass limit of the ICMF increases approximately linearly with the SFR surface density of the host galaxy \citep[e.g.][also see \citealt{kruijssen12b}]{johnson17}. As discussed in Section~\ref{sec:cfe}, this proportionality reflects that the maximum cluster mass increases with the gas pressure. Indeed, simple physical arguments can be made that the maximum cluster mass is proportional to the largest spatial scale that can achieve gravitational collapse (enhanced by pressure) before it is dispersed by centrifugal forces (enhanced by rotation). The resulting mass scale is referred to as the \citet{toomre64} mass, and naively its product with the SFE (setting a maximum stellar mass) and CFE (setting a maximum bound stellar mass) provides a natural upper limit for the ICMF \citep[e.g.][]{kruijssen14c}. However, in environments of both low gas pressure and low rotation, such as local-Universe galaxy disks (including the solar neighborhood), this model predicts upper mass limits that are too high compared to the observed masses of GCs. In these low-pressure environments, stellar feedback from the newborn massive stars is likely to disrupt the parent GMC before gravitational collapse can proceed to form high-mass clusters, resulting in a lower upper mass limit for the ICMF \citep{reinacampos17}.

The above physical picture provides context to the observed scaling between the upper mass limit of the ICMF and the SFR surface density, and additionally makes the concrete prediction that high-pressure environments with high rotation (such as galaxy centers) must have considerably lower upper mass limits than expected from the observed scaling relation. Indeed, such suppressed masses may be observed in the Central Molecular Zone of the Milky Way, where the most massive YMCs are just $\sim10^4~\msun$ \citep[e.g.][]{longmore14}. These masses are naturally reproduced by the rotation-limited model \citep{trujillogomez19}, whereas the scaling with the SFR surface density would predict $\sim10^6{-}10^7~\msun$ \citep{johnson17}. The natural prediction of this model is that the upper mass limit of the ICMF increases from $\sim10^4{-}10^5~\msun$ in the low-pressure environments of nearby galaxy disks to $\sim10^6{-}10^8~\msun$ in the high-pressure environments of high-redshift galaxies and galaxy mergers \citep[e.g.][]{adamo20b}. This adds a natural explanation to the observation that the formation of GC-like clusters is a rare event in the local Universe, but must necessarily have been more common at the formation redshifts of GCs.

The lower limit $M_{\rm min}$ of the double-Schechter ICMF is much more challenging to understand due to a lack of observational constraints. In external galaxies, the lower mass limit falls below the detection limit of the most sensitive surveys, which historically has led to several false positive detections of a lower cluster mass limit. As a result, the lower mass limit of the ICMF has mostly been constrained through theory and by indirect observational measurements. Observationally, the lower mass limit in the solar neighborhood seems to be of the order of $50{-}100~\msun$ \citep[e.g.][]{lada03,lamers05b}. In the formation environments of some GCs populations in dwarf galaxies, the lower mass limit must have been considerably higher than this, because there are not enough low-metallicity field stars to accommodate the destruction of a significant population of low-mass GCs \citep[e.g.][]{larsen12,larsen14}. Based on this observational result, \citet{trujillogomez19} proposed that the lower mass limit of the ICMF may be elevated at extremely high gas densities due to runaway hierarchical merging of forming clusters due to the inability of stellar feedback to halt the collapse process at the high gas pressures that are typical in these environments. In the dwarf galaxies considered, this would have resulted in a lower mass limit of $10^3{-}10^4~\msun$ \citep{trujillogomez19}, still more than an order of magnitude lower than the current typical GC mass, but also more than an order of magnitude higher than the lower mass limit in the solar neighborhood.

In summary, we see that GC formation might have been relatively common under the high-pressure conditions that are ubiquitous in high-redshift galaxies, due to a combination of a high CFE, a high ICMF upper mass limit, and potentially additionally a somewhat elevated ICMF lower mass limit. The typical gas pressures in these environments are $P/k\sim10^6{-}10^8~\kcmc$ \citep[e.g.][]{tacconi20}, which is a factor of $\sim10^3$ higher than the gas pressures in the solar neighborhood, but similar to the gas pressures observed in nearby galaxy mergers and galaxy centers \citep[e.g.][]{kruijssen13c}. In these environments, YMC formation is indeed observed to be common \citep[e.g.][]{krumholz19}.

\section{Star Cluster Destruction} \label{sec:destruction}
Once star clusters have formed, a wide variety of physical processes can lead to their destruction, implying that the origin of the GC population cannot be understood without considering these processes. The present-day GC population has a characteristic mass at $M\sim10^5~\msun$ \citep[e.g.][]{harris91}. To fully understand how this characteristic scale might have evolved from a power law-like double-Schechter ICMF at the time of their formation, it is necessary to describe how destruction processes have affected GCs over nearly a Hubble time of evolution. In this section, we provide a brief overview of these processes. We first focus on externally-driven cluster destruction processes, such as transient tidal perturbations due to interactions with the host galaxy or its substructure (`tidal shocks') and dynamical friction-driven inspiral into the host galaxy center. Subsequently, we discuss internally-driven cluster destruction processes, such as the tidal evaporation of star clusters due to two-body encounters between their constituent stars and black holes.

\subsection{Externally-Driven Cluster Destruction Processes} \label{sec:external}
Star clusters do not exist in isolation, but experience gravitational perturbations from their environment. It has been known for more than half a century that rapid changes in the tidal field experienced by a star cluster can lead to its disruption \citep[e.g.][]{spitzer58,ostriker72}. Such tidal perturbations, often referred to as `tidal shocks', inject kinetic energy into the cluster, causing its constituent stars to accelerate and thereby driving mass loss from the cluster \citep[e.g.][]{kundic95}. When estimating the magnitude of the resulting mass loss, it is important to distinguish two different regimes of shock duration. If the shock duration is short compared to the cluster crossing time, the shock is `impulsive' and the mass loss is maximal, because the internal motion of the stars does not have time to redistribute the injected energy. However, if the shock duration is long compared to the crossing time, the shock is `adiabatic' and the mass loss is reduced, because the energy is injected over the course of many orbital periods \citep[e.g.][]{spitzer87,weinberg94a,weinberg94b,weinberg94c}.

Initially, most research focused on the tidal shocks generated by macroscopic morphological elements of galaxies, such as galaxy disks and galaxy bulges \citep[e.g.][]{gnedin97,fall01}. However, these do not meaningfully amplify the dynamical mass loss from clusters beyond the mass loss driven by two-body relaxation (see Section~\ref{sec:internal}) due to their smooth potential and long durations compared to the crossing time, leading to the adiabatic damping of the shock \citep[e.g.][]{dinescu99,gnedin99c,baumgardt03,kruijssen09}. The maximum impact of tidal shocks increases with the density of the perturber \citep[e.g.][]{spitzer87}. As a result, the focus shifted to the substructure of galaxies as an important source of tidal shocks. \citet{gieles06,gieles07} showed that tidal shocks generated by spiral arms and especially by encounters with GMCs are able to drive significant dynamical mass loss from star clusters. This is particularly important in the context of the formation of GCs, because the high gas densities of their formation environments make them a natural source of tidal shocks \citep[e.g.][]{elmegreen10,kruijssen15b}.

\begin{figure}[t]
  \centering
  \includegraphics[width=\textwidth]{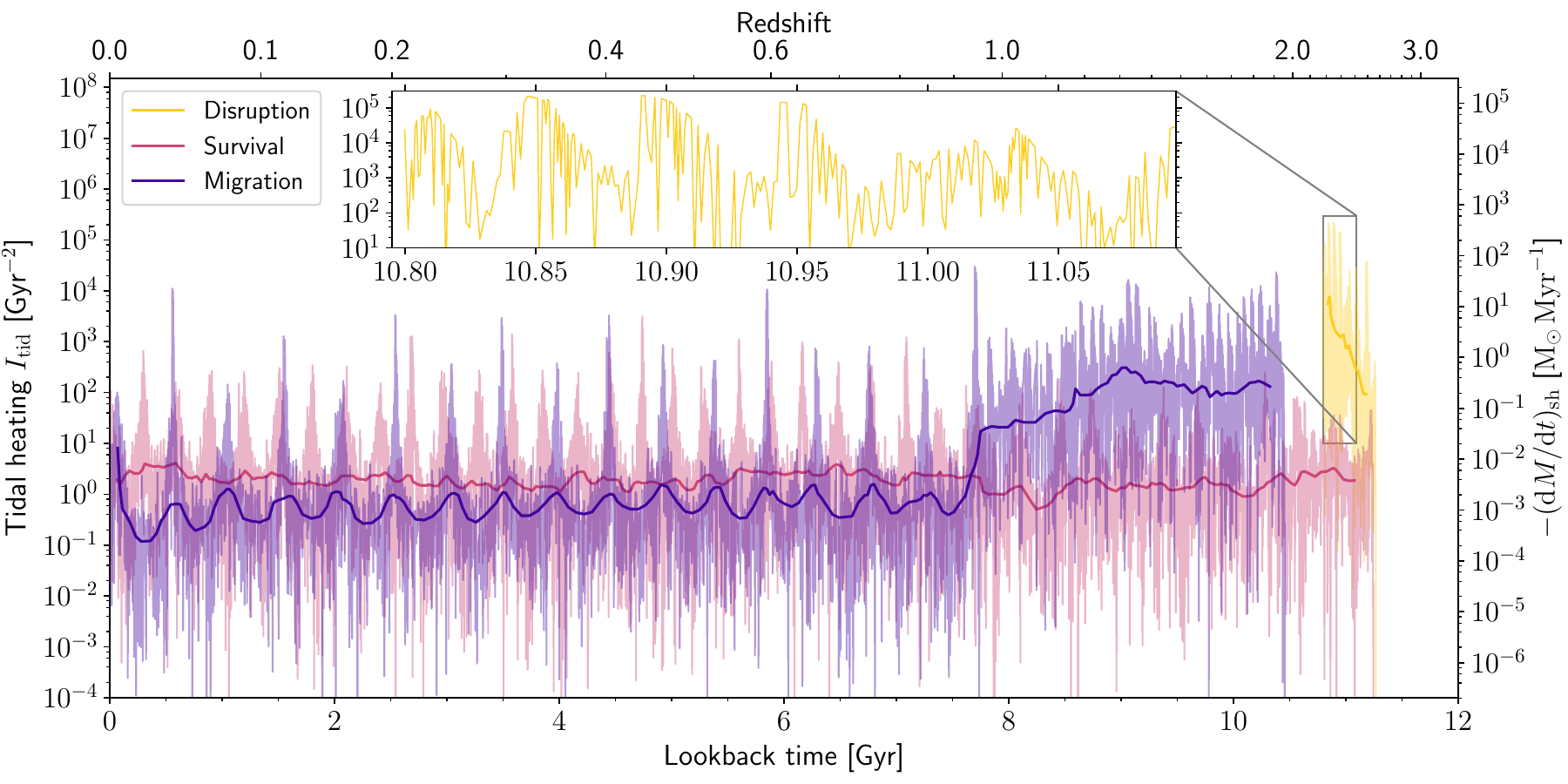}
  \caption{Illustration of the tidal heating history experienced by three different clusters with initial masses of 20,000~$\msun$ in one of the E-MOSAICS simulations of GC formation and evolution during galaxy formation \citep{pfeffer18}. The tidal heating parameter is calculated at a temporal resolution of 1~Myr, showing how rapid changes in the tidal field add kinetic energy to the clusters over the course of their evolution. The axis on the right-hand side converts the tidal heating into a mass loss rate according to \autoref{eq:tdis_sh}, assuming a cluster half-mass radius of 4~pc and shock intervals of 10~Myr. Each color represents a different cluster, with the thick lines showing a moving median. All three clusters meet different fates. In the `Disruption' case, the cluster forms in a high-pressure environment and experiences high tidal heating, leading to its destruction within 0.5~Gyr. In the `Survival' case, the cluster forms in a low-pressure environment and experiences low tidal heating, allowing it to survive until the present-day with a mass of 3000~$\msun$. In the `Migration' case, the cluster forms in a high-pressure environment, but at a lookback time of 7.5~Gyr a minor merger causes its migration into an environment of low tidal heating, allowing it to survive until the present-day with a mass of 3200~$\msun$. The oscillatory patterns in its subsequent tidal heating history reflect its motion on an eccentric orbit around the host galaxy. These three cases illustrate the uniqueness of the tidal heating history of each cluster, and the challenge of achieving a generalized description of tidal shock-driven cluster destruction. Taken from \citet{pfeffer18}, reproduced with permission.}
  \label{fig:heating_timesteps}
\end{figure}

In view of the great variety of (sub)structure capable of driving tidal shocks, it is necessary to formulate a generalized and process-agnostic description of shock-driven mass loss. \citet{prieto08} and \citet{kruijssen11} built on the extensive literature laying out tidal shock theory to identify the main proportionality for defining the cluster destruction timescale due to tidal shocks:
\be
\label{eq:tdis_sh}
t_{\rm sh}\equiv-\left(\frac{{\rm d}\ln{M}}{{\rm d}t}\right)_{\rm sh}^{-1} \propto \rho I_{\rm tid}^{-1}\Delta t ,
\ee
where $\rho$ is the cluster mass density, $I_{\rm tid}$ is the tidal heating parameter (defined below), and $\Delta t$ is the time since the previous shock. This scaling relation shows that tidal shock-driven cluster destruction is faster for clusters with lower densities, for environments with larger tidal heating parameters, and for shocks that occur more frequently. The tidal heating parameter is a measure of the energy injected into the cluster, and is obtained by integrating and then summing all components of the tidal tensor over the course of the shock. This approach enables the use of arbitrary tidal histories for modelling star cluster destruction. The natural challenge is that each tidal history is unique (illustrated in \autoref{fig:heating_timesteps}, also see e.g.\ \citealt{kruijssen11}, \citealt{li19}, and \citealt{meng22}), and designing a universal framework for tidal shock-driven cluster destruction models remains a challenge \citep[e.g.][]{webb19,webb24}.

Most importantly for GC formation, the destruction timescale due to encounters with substructure in the ISM can be simplified by assuming that the host galaxy disk resides in hydrostatic equilibrium and the cluster radius is constant, resulting in \citep[also see \citealt{elmegreen10}]{kruijssen15b}:
\be
\label{eq:tdis_sh2}
t_{\rm sh}=176~\myr~\left(\frac{f_\Sigma}{4}\right)^{-1}\left(\frac{\rho_{\rm ISM}}{\msun~\pc^{-3}}\right)^{-3/2}\left(\frac{M}{10^5~\msun}\right)\phi_{\rm ad}^{-1} ,
\ee
where $f_\Sigma$ is the surface density ratio between GMCs and the mean ISM surface density of the galaxy disk, $\rho_{\rm ISM}$ is the mean mid-plane gas mass density of the ISM, $M$ is the cluster mass, and $\phi_{\rm ad}$ is the `adiabatic correction' factor ($\phi_{\rm ad}\in[0,1]$) to reflect whether the shock is impulsive or adiabatic. The approximate magnitude of tidal shock-driven cluster destruction in \autoref{eq:tdis_sh2} reproduces the age distributions of young cluster populations in the local Universe, including the solar neighborhood \citep{lamers06b}, the Central Molecular Zone \citep{kruijssen14b}, and nearby galaxies \citep[e.g.][]{bastian12,miholics17}. These numbers imply that even relatively massive GC progenitors cannot have resided in their natal environment for much longer than a Gyr without being destroyed by tidal shocks. This is why the cluster migration process illustrated in Figure~\ref{fig:architecture} must have been a common occurrence in the early Universe, consistent with the elevated galaxy merger rates at high redshifts \citep[e.g.][]{kravtsov05,kruijssen15b}.

Even if a cluster escapes tidal shock-driven destruction by migrating out of the natal galaxy disk, it may be destroyed by dynamical friction-driven inspiral into the host galaxy center \citep[e.g.][]{tremaine75,goerdt06,antonini13,gnedin14,shao21}. As the ISM of galaxies gets denser towards the center, it is plausible that the cluster is destroyed by tidal shocks before reaching the nucleus. If the cluster is dense enough to survive, the inspiral timescale to the very center is given by \citep{kruijssen14c}:
\be
\label{eq:tdis_fric}
t_{\rm df}=2.5~\gyr~\left(\frac{V}{100~\kms}\right)\left(\frac{R}{1~\kpc}\right)^2\left(\frac{M}{10^6~\msun}\right)^{-1} ,
\ee
where $V$ is the circular velocity of the host galaxy at the orbit of the cluster, $R$ is the galactocentric radius, and $M$ is the cluster mass. This equation shows that dynamical friction proceeds faster for clusters with higher masses, lower circular velocities, and smaller galactocentric radii. It is much slower than tidal shock-driven cluster destruction, but it can proceed over much longer timescales, because it does not require the clusters to reside in the gaseous environnment of the host galaxy disk and proceeds long after cluster migration. In practice, dynamical friction affects GCs that either reside in the inner few kpc of their host galaxy, and/or have masses well above $\sim10^6~\msun$.

\subsection{Internally-Driven Cluster Destruction Processes} \label{sec:internal}
Once clusters have migrated to no longer permanently reside in the gaseous environment of their natal disk, the dominant destruction process is the tidal evaporation driven by internal two-body relaxation. The timescale for cluster destruction by tidal evaporation was long thought to be proportional to the half-mass relaxation time of the cluster, resulting in an evaporation-driven destruction time ($t_{\rm evap}\propto M^{1/2}r_{\rm h}^{3/2}$, with $r_{\rm h}$ the half-mass radius) that is proportional to the product of the crossing time $t_{\rm cr}\propto M^{-1/2}r_{\rm h}^{3/2}$ and the cluster mass $M$ \citep[e.g.][]{chernoff90,fall01,mclaughlin08}. However, the joint work of \citet{fukushige00} and \citet{baumgardt01} showed that stars do not escape instantaneously when they reach positive energies, but require a finite number of crossing times for their orbit to align with one of the Lagrange points within the host galaxy's tidal field and escape. This results in a destruction timescale due to evaporation with the proportionality \citep{baumgardt03}:
\be 
\label{eq:tdis_evap}
t_{\rm evap}\propto t_{\rm rh}^{3/4}t_{\rm cr}^{1/4}\propto\left(\frac{M}{\ln{\Lambda}}\right)^{3/4}t_{\rm cr}\rightarrow M^{0.6{-}0.7}\frac{V}{R} ,
\ee
where $t_{\rm rh}$ is the half-mass relaxation time, $t_{\rm cr}$ is the crossing time, $M$ is the cluster mass, $\ln{\Lambda}$ is the Coulomb logarithm, and $V/R$ is the ratio between the circular velocity of the host galaxy at the orbit of the cluster and its galactocentric radius (i.e.\ the orbital angular velocity).

The above proportionality of the evaporation-driven destruction timescale is largely insensitive to the structural properties of the clusters \citep{gieles08}, and as a result it is possible to describe the impact of evaporation on the GC population using only their masses and environment. At the solar galactocentric radius, \autoref{eq:tdis_evap} results in $t_{\rm evap}>30~\gyr$ for clusters with $M\gtrsim10^5~\msun$ \citep{baumgardt03}. As a result, evaporation only contributes to the destruction of relatively low-mass GCs within a Hubble time and together with its dependence on the galactocentric radius this means that evaporation cannot explain why the present-day GC population has a near-universal characteristic mass at $M\sim10^5~\msun$ \citep{vesperini03}. Recent work has shown that the presence of a large population of black holes in GCs could accelerate their evaporation-driven mass loss rates, possibly up to the rate needed to explain their present-day mass distribution \citep{gieles23}. Future work integrating these effects into GC population modelling in the context of galaxy formation and evolution is needed to place this theory on a firmer footing.

\subsection{How Much More Massive Were Globular Clusters at Birth?} \label{sec:mass}
Having discussed the destruction mechanisms that have affected the GC population over a Hubble time, it is possible to estimate how much more massive GCs might have been at birth. As discussed in Section~\ref{sec:external}, most of the GC mass loss likely took place in their natal environments, and models incorporating the formation and evolution of their host galaxies are needed to be able to quantify the mass loss experienced by surviving GCs. In recent years, such models have become available, enabling a quantitative assessment of how much more massive GCs might have been at birth.

As a natural result of the increase of the cluster disruption timescale with GC mass, both for tidal shocks and evaporation, the relative amount of mass loss decreases with the present-day GC mass. Therefore, GCs with a negligible mass were considerably more massive at birth (even with just evaporation-driven mass loss), whereas GCs with a present-day mass of $M\sim10^5~\msun$ experienced moderate mass loss. Quantitatively, theoretical studies estimate that the `typical' GC (with a current mass of $M\sim10^5~\msun$) was a factor of $2{-}4$ more massive at birth \citep[e.g.][]{kruijssen15b,reinacampos18,baumgardt19}. For GC masses greater than $10^5~\msun$, the relative degree of mass loss was similar, because stellar evolution on its own causes a cluster to lose nearly 50\% of its mass over a Hubble time \citep[e.g.][]{lamers05b}. The minimum initial mass needed for survival until the present-day was plausibly $10^5~\msun$ for typical tidal histories \citep[e.g.][]{kruijssen15b,choksi18,reinacampos18}, implying that the current lowest-mass GCs (with masses of just a few $100~\msun$) were nearly a factor of $10^3$ more massive at birth. While the fractional mass loss thus turns out to be a steeply decreasing function of the present-day GC mass, typical GCs were only a factor of a few more massive at birth than they are today.

\section{Globular Cluster Formation and Destruction in the Context of Their Host Galaxies} \label{sec:galaxies}
Having reviewed the key physical mechanisms governing the demographics of the GC population at birth and thereafter, we now consider how these mechanisms manifest themselves in the context of galaxy formation and evolution. Observations of GC populations in nearby galaxies provide a wealth of demographic information that can be used to infer the conditions under which GCs formed. Until recently, it was a challenge to make this connection explicitly, but the development of comprehensive GC formation and evolution models within galaxy formation simulations has provided quantitative insight into the drivers behind the main observables used to describe GC demographics. In Table~\ref{tab:observables}, we summarize the main observational properties of GC populations and their physical origins. The ability of modern GC formation and destruction models to reproduce such a broad range of observed GC demographics reinforces the idea that GCs formed according to the same physical mechanisms as the YMCs that are still observed to form in the local Universe.

\subsection{Relations between the Number of Globular Clusters and the Host Galaxy Properties}
There exists a well-known relation between the mass of a galaxy's GC population (or its number of GCs) and the mass of its host dark matter halo, known as the GC system mass-halo mass relation. This relation dates back to the pioneering work of \citet{blakeslee97}, who found that the number of GCs per unit X-ray temperature in clusters of galaxies (which is a tracer of the halo mass) is constant across 19 Abell clusters. This first indication that the number of GCs is proportional to the galaxy halo mass has since been expanded across more than six orders of magnitude in halo mass ($M_{\rm halo}\sim10^8{-}10^{14}~\msun$) and a variety of galaxy types and environments \citep[e.g.][]{peng08,durrell14,hudson14,harris17,forbes18b,burkert20,eadie22,chen23}. A natural consequence of the relation between the number of GCs and the halo mass is that the specific frequency of GCs, which is the ratio between the number of GCs and the luminosity or stellar mass of the host galaxy \citep[e.g.][]{harris81}, also increases towards more massive galaxies for galaxies more massive than the Milky Way \citep[e.g.][]{peng08}. However, for lower-mass galaxies, the specific frequency increases too, resulting in a U-shaped relation, albeit with considerable scatter \citep[e.g.][]{peng08}. As a result, the upturn at high galaxy masses is interpreted as the result of the accretion of dwarf galaxies with high specific frequencies. In the subclass of ultra-diffuse galaxies (UDGs), the specific frequency of GCs seems to be higher than in normal galaxies \citep[e.g.][]{lim18,vandokkum18b,mueller20,danieli22}.

The tempting interpretation of the above relations, which has been quite popular across the GC population modelling literature for about a decade, is that GCs are (possibly randomly) seeded in dark matter halos, and that the GC system mass-halo mass relation reflects the hierarchical assembly of galaxies and the growth of their dark matter halos, which through the central limit theorem linearizes the relation between the number of GCs and the halo mass \citep[e.g.][]{boylankolchin17,elbadry19,valenzuela21}. The strong prediction of this interpretation is that the GC system mass-halo mass relation should develop considerable scatter at the lowest halo masses, because there the number of GCs is dominated by the seeding of GCs in halos. While there is no doubt that hierarchical galaxy assembly helps linearizing the GC system mass-halo mass relation, it is hardly physical to attribute the relation entirely to sampling statistics. GCs form from gas, not from dark matter. The natural correlation should be between the number of GCs and the integral of the host galaxy's SFR, with second-order trends plausibly driven by the environmental variation of the CFE and the high-mass truncation of the double-Schechter ICMF.

\begin{table}[t]
  \TBL{\caption{Summary of the main observational properties of GC populations and their physical origins. The favored physical explanation is listed in each instance; for some, alternatives are provided in the text. All favored explanations fit within the self-consistent GC formation and destruction framework of \autoref{fig:architecture}. Each explanation is accompanied by references supporting (aspects of) the explanation.}\label{tab:observables}}
  {\begin{tabular*}{\textwidth}{@{\extracolsep{\fill}}@{}lll@{}}
  \toprule
  \multicolumn{1}{@{}l}{\TCH{Observable}} &
  \multicolumn{1}{l}{\TCH{Favored Physical Explanation}} &
  \multicolumn{1}{l}{\TCH{References}}\\
  \colrule
  GC system mass-halo mass relation & GC destruction stronger in high-baryon halos + hierarchical assembly & 1, 2, 3, 4 \\
  Specific frequency & GC destruction stronger in high-mass galaxies + dwarf galaxy accretion & 1, 2, 4, 5, 6 \\
  GC mass function & Tidal shock-driven destruction at low masses, saturating at high masses & 1, 7, 8 \\
  GC upper truncation mass & ISM pressure-dependent maximum cluster mass + dynamical friction & 9, 10, 11 \\
  GC metallicity distribution & Host galaxy assembly history + cluster destruction + cluster formation & 12, 13, 14 \\
  Blue tilt \& GC metallicity floor & Galaxy mass-metallicity relation + minimum mass for GC formation & 15, 16 \\
  GC age(-metallicity) distribution & Host galaxy star formation and assembly history & 17, 18, 19, 20 \\
  GC spatio-kinematic distribution & Host galaxy assembly history & 21, 22, 23 \\
  \botrule
  \end{tabular*}}{%
  \begin{tablenotes}
  \footnotetext{References: (1) \citet{kruijssen15b}, (2) \citet{choksi19}, (3) \citet{elbadry19}, (4) \citet{bastian20}, (5) \citet{peng08}, (6) \citet{lamers17}, (7) \citet{elmegreen10}, (8) \citet{reinacampos22}, (9) \citet{gieles09}, (10) \citet{pfeffer18}, (11) \citet{hughes22}, (12) \citet{muratov10}, (13) \citet{tonini13}, (14) \citet{pfeffer23}, (15) \citet{usher18}, (16) \citet{kruijssen19c}, (17) \citet{leaman13}, (18) \citet{kruijssen19d}, (19) \citet{li19}, (20) \citet{delucia24}, (21) \citet{trujillogomez21}, (22) \citet{chen22}, (23) \citet{reinacampos22b}.}
  \end{tablenotes}
  }%
\end{table}

A key insight into the physical drivers of relations between the number of GCs and the host galaxy's properties is that the specific frequency is a monotonically decreasing function of metallicity when dividing the stars and GCs in a galaxy into narrow metallicity bins \citep[also see \citealt{harris02}]{lamers17}. This is an incredibly meaningful observation -- to first order, a GC's metallicity traces the mass of its host galaxy at the time of its formation \citep[much more so than its formation redshift, see e.g.][]{ma16}. Higher-mass galaxies have deeper gravitational potential wells, resulting in higher gas pressures, and correspondingly more rapid tidal shock-driven cluster destruction. In other words, the decrease of the specific frequency with metallicity encodes how the cluster destruction rate has affected the GC population as a function of the host galaxy's mass. The specific frequency does not reflect a formation efficiency \citep[as was originally thought, see e.g.][]{peng08,harris15}, because the dynamic range of expected formation efficiencies \citep[e.g.][]{adamo20b} is too small to explain the observed dynamic range of specific frequency as a function of metallicity. Instead, it reflects a destruction efficiency \citep{lamers17}. The slope of this trend is deceptively similar (but opposite) to the slope of the galaxy stellar mass increasing with the halo mass \citep[e.g.][]{moster13}. As a result, the number of surviving GCs at birth (but not necessarily the number of all potential GCs that are born) scales with the halo mass as $N_{\rm GC}/M_{\rm halo}\propto M_{\rm star}^{0.1}$ \citep{kruijssen15b}. Hierarchical galaxy assembly is comfortably capable of linearizing this birth relation to the observed GC system mass-halo mass relation \citep[e.g.][]{elbadry19}.

\begin{figure}[t]
  \centering
  \includegraphics[width=0.55\textwidth]{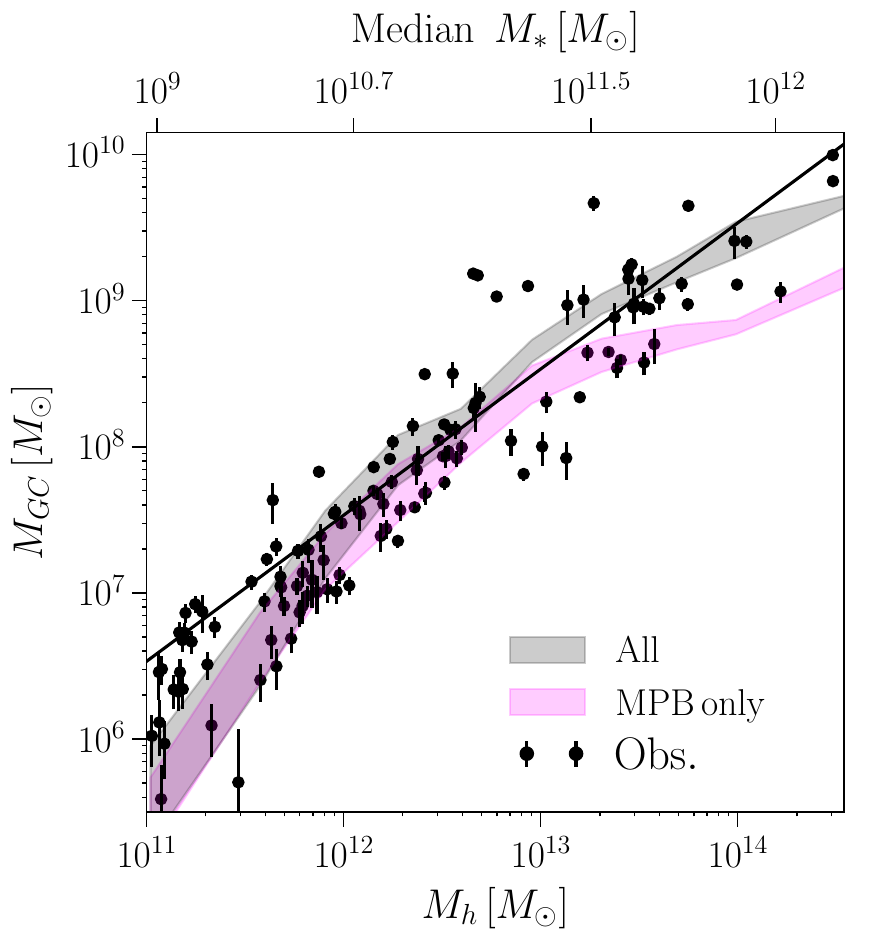}
  \caption{Relation between the GC system mass and the host galaxy halo (bottom axis) or stellar (top axis) mass. Observations are shown as the black symbols with error bars, whereas colored bands show predictions of a model that assumes cluster formation takes place in proportion to a galaxy's gas mass at times of elevated galaxy growth rates, and adopts a simple parameterization for a cluster mass-dependent destruction rate. The bands refer to model predictions when only including GCs that formed in the main progenitor branch of the host galaxy (`MPB'), and when also including GCs that formed in accreted satellite galaxies (`All'). The figure illustrates that the accretion of satellite galaxies with high specific frequencies is necessary to reproduce the scaling at high galaxy masses. Taken from \citet{choksi19}, reproduced with permission.}
  \label{fig:choksi19}
\end{figure}

Galaxy formation simulations and theoretical models that include a physical description for GC formation and destruction indeed reproduce the observed scaling relations governing the number of GCs in most galaxies \citep[see \autoref{fig:choksi19} and e.g.][]{kruijssen15b,choksi19,bastian20,delucia24}. The scatter on these modelled relations does increase towards lower galaxy masses as expected in the simple hierarchical assembly models, but insufficiently to explain the observed scaling relations entirely as a statistical effect \citep{bastian20}. The galaxy mass dependence of GC destruction seems to be key. This also suggests that the elevated specific frequency of GCs in UDGs may be explained by cutting short the epoch over which GCs can be efficiently destroyed by tidal shocks, for instance by rapid GC migration due to galaxy merging \citep[e.g.][]{trujillogomez21b,trujillogomez22,vandokkum22,pfeffer24b}.

\subsection{The GC Mass and Luminosity Functions}
Within a galaxy, the distribution of GC masses (often referred to as the GC mass function or GCMF) or luminosities is a key observable that differs fundamentally from the power law-like double-Schechter ICMF by virtue of its peaked shape, with a characteristic mass at $M\sim10^5~\msun$ \citep[e.g.][]{harris91,jordan07,caldwell16,baumgardt18}. Many decades of literature have revolved around the physical drivers of this shape and its stark contrast to the ICMF. While initial work reasoned it might represent the initial mass distribution of GCs at birth \citep[e.g.][]{parmentier05}, the more popular interpretation quickly shifted to the idea that the characteristic mass emerged due to the destruction of lower-mass GCs \citep[e.g.][]{fall01}. The destruction mechanisms remained hotly debated, even until the present day. Initial work focused on the evaporation-driven destruction of GCs \citep[e.g.][]{fall01,vesperini01,mclaughlin08,kruijssen09b}, but the strong dependence of this mechanism on the galactocentric radius (see \autoref{eq:tdis_evap}) results in a characteristic mass scale that decreases with distance to the host galaxy, in contrast to observations \citep[e.g.][]{vesperini03}.

Most recently, the focus has shifted to the tidal shock-driven destruction of GCs \citep[e.g.][]{elmegreen10}. This picture is tempting, because tidal shocks are much more effective at cluster destruction than evaporation, and the necessary cluster destruction occurs while the clusters still reside in their natal galaxy disks, before they migrate to their present-day locations, thus erasing any variation of the GCMF with galactocentric radius \citep{kruijssen15b}. Galaxy formation models that include a physical description for GC formation and destruction must also include the main destruction agent to be able to reproduce the observed GCMF. Therefore, simulations without a cold ISM result in a GCMF with a characteristic mass scale that is considerably lower than observed \citep{pfeffer18,kruijssen19d}. The latest generation of simulations that does include a cold ISM does reproduce the observed GCMF \citep{reinacampos22}, showing that its peaked shape is in place within a few Gyr after GC formation due to tidal shock-driven cluster destruction, before tidal evaporation has had time to modify the GCMF.

The increase of the ISM pressure and the cluster destruction rate with host galaxy mass results in a GCMF characteristic mass that very weakly increases with the host galaxy mass \citep{kruijssen15b}, in agreement with observations \citep{jordan07}. However, the magnitude of this trend is extremely minor, to the point that the characteristic mass of the GCMF is commonly used as a distance indicator \citep[e.g.][]{rejkuba12}. How likely is it that all galaxies hosted such a similar degree of GC destruction that the GCMF characteristic mass is universal? Part of that question may be answered by the fact that the presence of a high-pressure (and highly destructive) ISM is a prerequisite for GC formation \citep{kruijssen15b}, plausibly resulting in similar degrees of cluster destruction. However, perhaps more important is the fact that the effect of cluster destruction on the GCMF can saturate. This is caused by the exponential truncation at the high-mass end in \autoref{eq:icmf}, which means that the increase of the characteristic mass due to the ongoing destruction of low-mass clusters saturates when the characteristic mass reaches approximately $10\%$ of the truncation mass \citep{gieles09}. This saturation effect implies that cluster destruction does not need to have had a universal impact to explain the universal characteristic mass of the GCMF, but it must have reached a minimum integrated effect.

Interestingly, the exponential truncation mass at the high-mass end of the GCMF also weakly increases with galaxy mass \citep[e.g.][]{jordan07}. This is thought to be driven by the increase of the ISM pressure towards more massive galaxies and the weaker effect of dynamical friction in these systems \citep[e.g.][]{pfeffer18,hughes22}. It is not inconceivable that the weak increase of the characteristic mass of the GCMF with galaxy mass is related to the weak increase of the truncation mass with galaxy mass, simply representing its saturation point.

It is somewhat unsatisfactory that tidal shocks seem to play such a major role in explaining the present-day demographics of the GC population. Their effectiveness depends strongly on poorly-constrained quantities such as the cluster density (see \autoref{eq:tdis_sh}; initial GC radii cannot be reconstructed from their present-day properties, see \citealt{gieles11}), and the infinite number of unique tidal histories implies that tidal shock-driven cluster destruction is a highly stochastic (and plausibly non-deterministic) process. This situation has led to research into ways in which the effectiveness of other cluster destruction mechanisms may be enhanced. The discovery of significant populations of black holes in GCs \citep[e.g.][]{strader12,giesers18} has prompted the hypothesis that their presence may accelerate the evaporation-driven mass loss of GCs \citep{gieles23}. This mechanism is quite sensitive to the ability of GCs to retain their black holes, and it is still unclear to what extent this mechanism may explain the full range of observed GC demographics. Therefore, further research (up to a similar level of scrutiny as tidal shock-driven cluster destruction) is required to understand how the full suite of cluster destruction mechanisms may have affected the GC population. As is often the case in astrophysics, the field should not be drawn into a false dichotomy, but realize that multiple mechanisms may be at play and interact in complex ways.

\subsection{The GC Metallicity Distribution}
The chemical composition of GCs (generally expressed in terms of its iron content $\feh$) is a fundamental property of the GC population, because it reflects the mass of its host galaxy at the time of its formation (see above). Historically, it has been probed through the distribution of GC broadband colors \citep[e.g.][]{forbes97,larsen01,peng06,chiessantos12}. In many galaxies (including the Milky Way), the GC metallicity distribution is well-described by a bimodal distribution, separating into metal-poor (or `blue') and metal-rich (or `red') GCs \citep[e.g.][]{harris96,larsen01,brodie12}. Both of these populations show kinematic and spatial differences \citep[e.g.][]{zinn85,dinescu99}, with the metal-rich GCs being associated with the host galaxy centroid and the metal-poor GCs being scattered throughout the galaxy halo.

Historically, these two populations have led to GC formation theories that invoke different formation mechanisms \citep[e.g.][]{forbes97,beasley02}. However, the discovery that bimodal GC metallicity distributions are not necessarily a universal feature of galaxies and instead galaxies exhibit a wide range of differently shaped metallicity distributions \citep[e.g.][]{usher12,harris17b} led to the realization that the metallicity distribution of GCs might emerge as a result of GC formation and destruction processes in the context of the assembly history of their host galaxy. \citet[also see \citealt{muratov10}]{tonini13} used a simple model tagging GCs to dark matter halos in a large cosmological simulation to study the origin of the GC metallicity distribution, proposing that galaxy assembly shapes the GC metallicity distribution by adding metal-poor GCs from accreted dwarf galaxies to a predominantly metal-rich population that formed in the main progenitor of the host galaxy. Depending on the assembly history, this may naturally result in a bimodal distribution \citep[an idea dating back to][]{cote98}. \citet[also see \citealt{li19}]{pfeffer23} showed that a more physically-motivated model for GC formation and destruction reveals two regimes for generating GC metallicity bimodality. In galaxies with masses below $10^{10}~\msun$, metallicity bimodality exists at the time of cluster formation, and thus must reflect non-monotonic variations in the CFE and ICMF upper truncation mass. In galaxies with masses above $10^{10}~\msun$, metallicity bimodality typically only manifests itself at $z=0$, after considerable evolution. This bimodality is uncorrelated with typical metallicities of in-situ and ex-situ GCs, and instead is related to cluster destruction \citep{pfeffer23}. As a result, nearly half of the galaxies with masses above $10^{10.5}~\msun$ exhibit a bimodal GC metallicity distribution, in quantitative agreement with observations \citep{usher12}. Taken together, the above findings show that the metallicity distribution of GCs is an emergent property of the interplay between GC formation, destruction, and the assembly history of their host galaxy, rather than a fundamental tracer of GC formation physics.

The low-metallicity end of the GC metallicity distribution exhibits a sharp cutoff, often referred to as the GC metallicity floor \citep[e.g.][]{beasley19}. This cutoff is thought to reflect the minimum mass that a galaxy needs to have in order to form a GC massive enough to survive until the present day, through the galaxy mass-metallicity relation at the GC formation redshift \citep[also see \citealt{choksi18}]{kruijssen19c}. This interpretation also implies that the GC metallicity floor should increase for more massive GCs, which naturally explains the observation that the most massive and metal-poor GCs exhibit a mass-metallicity relation, with more massive GCs being less metal-poor \citep[the `blue tilt', see e.g.][]{harris06,mieske10}. More complex explanations have also been offered, wherein the blue tilt results from the chemical self-enrichment of GCs during their formation \citep[e.g.][]{bailin09} or the fact that low-metallicity galaxies do not reach the high ISM pressures needed to form massive GCs \citep[e.g.][]{usher18}. The latter explanation is subtly different from the simple mass budget interpretation of \citet{kruijssen19c}, because it focuses on the dependence of the maximum GC mass on the conditions in the ISM, as a function of the host galaxy mass. The self-enrichment explanation is unlikely to apply, as the supernova ejecta that would need to be retained by GCs to generate the necessary self-enrichment have velocities well in excess of the escape velocity of GCs \citep[e.g.][]{bastian18}. In recent years, the debris of destroyed clusters with metallicities below the metallicity floor have been discovered \citep[e.g.][]{wan20,martin22}, supporting the idea that galaxies with metallicities below the metallicity floor were incapable of forming GCs with masses high enough to survive until the present day. However, the discovery of EXT8, a GC with a mass of $\sim10^6~\msun$ and a metallicity of $\feh\sim-2.9$ residing in the nearby Andromeda galaxy \citep{larsen20}, shows that exceptions do exist. Uncovering the physical processes that led to the formation of EXT8 is an important frontier in current GC formation research.

\subsection{The GC Age(-Metallicity) Distribution}
\begin{figure}[t]
  \centering
  \includegraphics[width=0.9\textwidth]{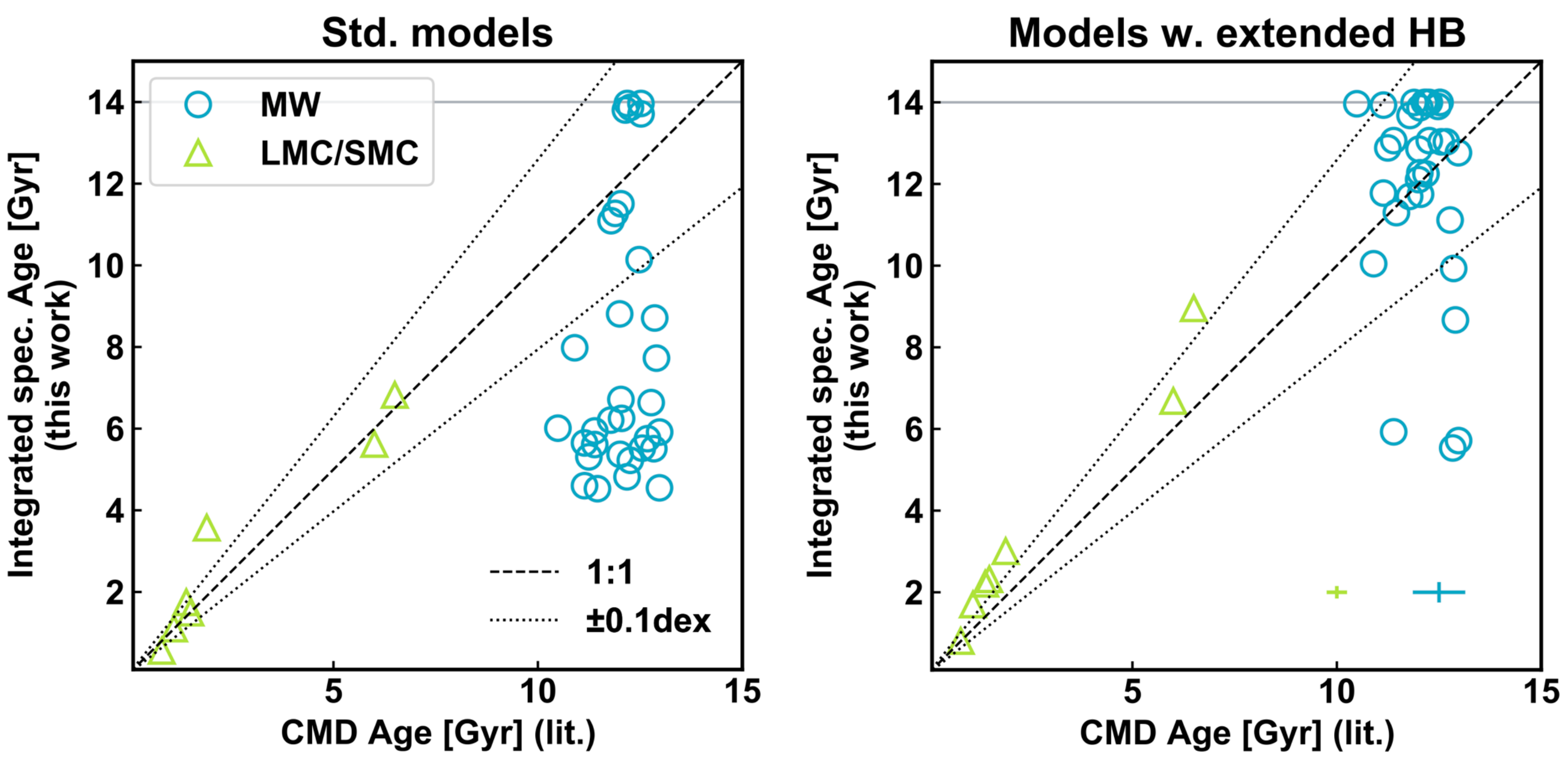}
  \caption{Comparison of GC ages measured using integrated light to the ages obtained from the color-magnitude diagrams, when using standard stellar population synthesis models (left panel) and when using models that include an extended (hot) horizontal branch (HB) population (right panel). Without modelling the extended HB population, the integrated light ages of GCs are systematically younger than the color-magnitude diagram ages. agreeing to within 25\% only in 28\% of the cases. With a more appropriate treatment of hot HB stars \citep{cabreraziri22}, the ages obtained from integrated light are in good agreement with the color-magnitude diagram ages, agreeing to within 25\% in 81\% of the cases. This improvement enables the use of ages in extragalactic GC studies to help distinguish between different GC formation theories and constrain the assembly histories of the host galaxies. Taken from \citet{cabreraziri22}, reproduced with permission.}
  \label{fig:cabreraziri22}
\end{figure}

The literature focusing on GC formation modelling has long argued that accurate measurements of GC ages represent the Rosetta stone for distinguishing between current GC formation theories \citep[e.g.][]{leaman13,li14,li19,forbes18}, because many of these models differ by the predicted cosmic epoch of peak GC formation (see Section~\ref{sec:history}). Unfortunately, accurate ($\sigma(\tau)<1~\gyr$) age measurements are extremely challenging and are restricted to the GC population of the Milky Way \citep[e.g.][]{forbes10,dotter10,dotter11,vandenberg13}. Only recently, the first age measurements have been made for the GC populations of external galaxies using integrated light \citep{usher19}. While these are still characterized by considerable uncertainties, they reveal major differences to the age distribution of Galactic GCs, for instance by extending down to younger ages than in the Milky Way. Additionally, the median ages of the GCs in these galaxies correlates with the median age of their field stars, which is consistent with the idea that GC formation is a natural consequence of high-pressure star formation \citep{usher19}. Major advances in stellar population modelling have since considerably improved the accuracy of ages obtained from integrated light measurements \citep[see \autoref{fig:cabreraziri22} and ][]{cabreraziri22}, so that GC ages might soon become a commonly used demographic for characterizing extragalactic GC systems.

The characterization of extragalactic age distributions has come at a convenient time, as the latest generation of GC formation models predict considerable variance in the age distributions of GCs between different galaxies \citep[e.g.][]{kruijssen19d,li19,delucia24}. Across these models, the GC age distribution is a strong function of the host galaxy mass and of its assembly history, with more massive galaxies and galaxies that assembled earlier having older GCs on average. In the context of these model predictions, the Milky Way GC population appears to be older than average, which may be the result of the Milky Way's relatively rapid assembly history \citep{kruijssen19e}. The galaxy mass dependence of the GC age distribution is best illustrated in age-metallicity space, where lower metallicity GCs extend to younger ages than higher metallicity GCs, indicative of the slower assembly of lower-mass galaxies, which continue to form GCs until later times than more massive galaxies \citep[e.g.][]{forbes10,leaman13,kruijssen19d,horta21b}. This means that the GC age(-metallicity) distribution is a strong function of the host galaxy's star formation and assembly history.

\subsection{GC Spatio-Kinematic Distribution}
The spatial and kinematic distributions of GCs have historically been used to shed light on their origins. The spatial distribution of metal-rich GCs is typically found to trace the host galaxy surface brightness profile, with metal-poor GCs generally forming a spatially more extended component \citep[e.g.][]{zinn85,rhode04,kartha14}. The spatial extent of the total GC population scales with the size of the host galaxy and its dark matter halo, and the GCs can be reasonably well-described with a one- or two-component power law or \citet{sersic63,sersic68} profile \citep[e.g.][]{rhode04,kartha14,hudson18}. The kinematics of the metal-rich GCs tend to be more rotationally supported, whereas the kinematics of the metal-poor GCs are more radially anisotropic \citep[e.g.][]{dinescu99,pota13}.

Key aspects of these observed spatial and kinematic properties are explained by considering the assembly of the GC population as a byproduct of the galaxy assembly process. The simultaneous hierarchical growth of galaxies and their GC populations naturally results in a correlation between the extent of the GC system and the size of the host galaxy. Indeed, modern GC formation models have no trouble reproducing observed spatial distributions of GCs \citep[e.g.][]{li19,reinacampos22,reinacampos22b}. The kinematics follow naturally too -- if metal-rich GCs are largely associated with the main progenitor of the host galaxy, they will have been only weakly affected by galaxy merger-induced migration and are likely to have retained some of their initial angular momentum \citep[e.g.][]{trujillogomez21}. By contrast, metal-poor GCs that formed in satellite galaxies will have inherited orbital properties based on the accretion orbit of their natal galaxy \citep[e.g.][]{hughes19,massari19}. While it seems that hierarchical galaxy assembly naturally leads to the observed spatial and kinematic distributions of GCs, it remains to be assessed whether the details of these distributions can be understood, as well as any trends with the host galaxy properties. These are important avenues for future work.

\section{Current Frontiers in Globular Cluster Formation and Concluding Perspectives} \label{sec:frontiers}
The major advances of the last decade have left GC formation research at a point where a comprehensive theory of GC formation and evolution within the context of galaxy formation is within reach for the first time. The current frontiers are focused on further expanding and utilizing our understanding of GC formation. We conclude this review with a short discussion of these frontiers and a perspective on the questions that will need to be addressed over the coming decade.

\subsection{Trinity of Globular Cluster Formation Diagnostics} \label{sec:trinity}
Historically, GC formation theories were mostly based on the properties that GC populations are observed to have at the present-day (see Section~\ref{sec:galaxies}), attempting to rewind the impact of cluster destruction. This is an intrinsically problematic approach for the simple reason that destruction erases a considerable part of the initial conditions that led to the GC population. A broad range of initial conditions might result in aspects of the present-day GC population that are similar, but the current hypothesis (or hope) is that very few of these initial conditions may reproduce all observables described above.

Thankfully, using the present-day properties of the GC population to infer their formation physics is not the only approach. The discovery of YMC formation in the nearby Universe provides a high-resolution perspective on the formation of GC-like clusters (see Section~\ref{sec:history}), which might end up becoming GCs themselves if the evolution of their environment enables their survival. This perspective allows testing our understanding of the physics of GC formation and evolution in much more detail than possible with the present-day GC population alone. Similarly, GC ages provide a reasonable idea of when most GCs might have formed. While extragalactic GC age measurements are starting to reveal a plurality of GC formation histories \citep{usher19}, it is clear that GC formation was more common at higher redshifts than at the present day. Characterizing the physical conditions under which GCs might have formed in high-redshift galaxies is therefore an important frontier in our quest to understand the formation of GCs. With the unprecedented sensitivity and resolution of telescopes such as the James Webb Space Telescope (JWST), it is becoming possible to perform direct observations of not just the environments of GC formation, but also of proto-GCs themselves (see Section~\ref{sec:jwst}). Of course, the question remains whether the observed proto-GCs will be destroyed, or if they can survive to become GCs at the present day. Only with statistically meaningful surveys of (proto)-GCs it is possible to answer this question, but observatories like the upcoming Extremely Large Telescope (ELT) will play a central role in starting to provide these across the necessary redshift range, from GC formation to the present day.

These developments imply that it is no longer sufficient to test a GC formation theory against all observables describing the present-day GC population. Instead, a theory must be tested against the three key diagnostics of GC formation: (1) the physics of YMC formation observed in the local Universe, (2) the conditions of GC formation observed at high redshifts, and (3) the demographics of (proto)-GCs from high redshift until the present day. This greatly expands the predictive power of GC formation theories, because it eliminates degeneracies between different models and tests our emerging physical picture against the full chronology of GC formation and evolution.

Out of these three key diagnostics, the first has been explored in detail over the past two decades, and the local-Universe part of the third diagnostic has historically formed the foundation of GC formation research. The big frontier is represented by the intermediate- and high-redshift diagnostics, which only gained traction over the past decade and will undergo a revolution with the current and upcoming generation of telescopes.

\subsection{Direct Observations of Globular Cluster Formation and Evolution at High and Intermediate Redshifts} \label{sec:jwst}
The most direct way to study the formation of GCs is to directly observe the formation of star clusters in the distant Universe. This is a highly challenging endeavor due to the extreme demands on sensitivity and angular resolution, where the latter is not aided by needing to make the observations at high redshifts. The first observations of YMC formation in the early Universe were made with HST by targeting strongly gravitationally lensed galaxies across the redshift range $z=2{-}6$ \citep[e.g.][]{johnson17b,vanzella17,vanzella17b,vanzella19}. These pioneering studies paved the way for routine observations of YMCs (in this context often referred to as `proto-GCs') out to even larger redshifts, now spanning $z=1{-}10$ \citep[e.g.][]{mowla22,mowla24,vanzella22b,vanzella23,claeyssens23,adamo24,fujimoto24}. Thanks to the combination of JWST's sensitivity and resolution with strong gravitational lensing, some of these proto-GCs can be resolved down to a few pc, confirming that many of these objects are indeed compact clusters and might be precursors of the GCs that we observe at the present day.

One of the biggest advantages of direct observations at the time of GC formation is that the younger ages of proto-GCs ($\tau\lesssim1~\gyr$) makes age measurements considerably less uncertain than for present-day (extragalactic) GC populations (with uncertainties $>1~\gyr$!). This means that the ability to distinguish GC formation theories based on the age distribution of proto-GCs will be a powerful diagnostic in the coming decade. The realization of this potential will rely critically on the ability to obtain statistically representative samples. Unfortunately, this remains challenging even with JWST, and sample definition and completeness are complex, as the observations depend strongly on the source plane reconstruction of the lensed background galaxies. Most of the current work has been focusing on the properties of the brightest proto-GCs, finding targets scattered across the mass range expected for the formation of GC progenitors \citep[$10^5{-}10^8~\msun$, e.g.][]{mowla22,mowla24,vanzella22,vanzella23,claeyssens23,adamo24,fujimoto24}. It is important to remember that the host galaxies have low masses at these redshifts, across the range $10^6{-}10^9~\msun$. This means that the proto-GCs may emit a large fraction of the galaxy light in the rest-frame optical, ranging from $1{-}50\%$, drawing a striking parallel with the expected dynamic range of the CFE in these galaxies \citep[about $5{-}50\%$, see e.g.][]{kruijssen12d,pfeffer24}.

Galaxy formation models that include GC formation and evolution have attempted to predict the rather sparsely sampled demographics of proto-GCs that can currently be obtained, so far with surprising success \citep[e.g.][]{pfeffer19,pfeffer24}. Despite the fact that no empirical input from proto-GC demographics was available at the time these simulations were performed, they reproduce the observed rest-frame ultraviolet luminosity function of proto-GCs \citep{bouwens21}, as well as the optical luminosities, masses, and ages of the brightest proto-GCs, the relation between their luminosities and the host galaxy's star formation rate, and the fraction of the host galaxy light that is emitted by proto-GCs \citep{pfeffer24}. Once statistically representative observational samples of proto-GCs are available, they may also enable testing the prediction that the peak of cosmic proto-GC formation precedes the peak of cosmic star formation by $\sim2~\gyr$ \citep[e.g.][]{choksi19b,reinacampos19,joschko24}. The key question remains which of these proto-GCs will actually survive until the present day and become actual GCs (Chevance et al.\ in prep.). Current models predict that the peak of \textit{surviving} GC formation may precede the peak of cosmic star formation by less than $\sim1~\gyr$ \citep[e.g.][]{choksi19b,joschko24}.

\begin{figure}[t]
  \centering
  \includegraphics[width=0.97\textwidth]{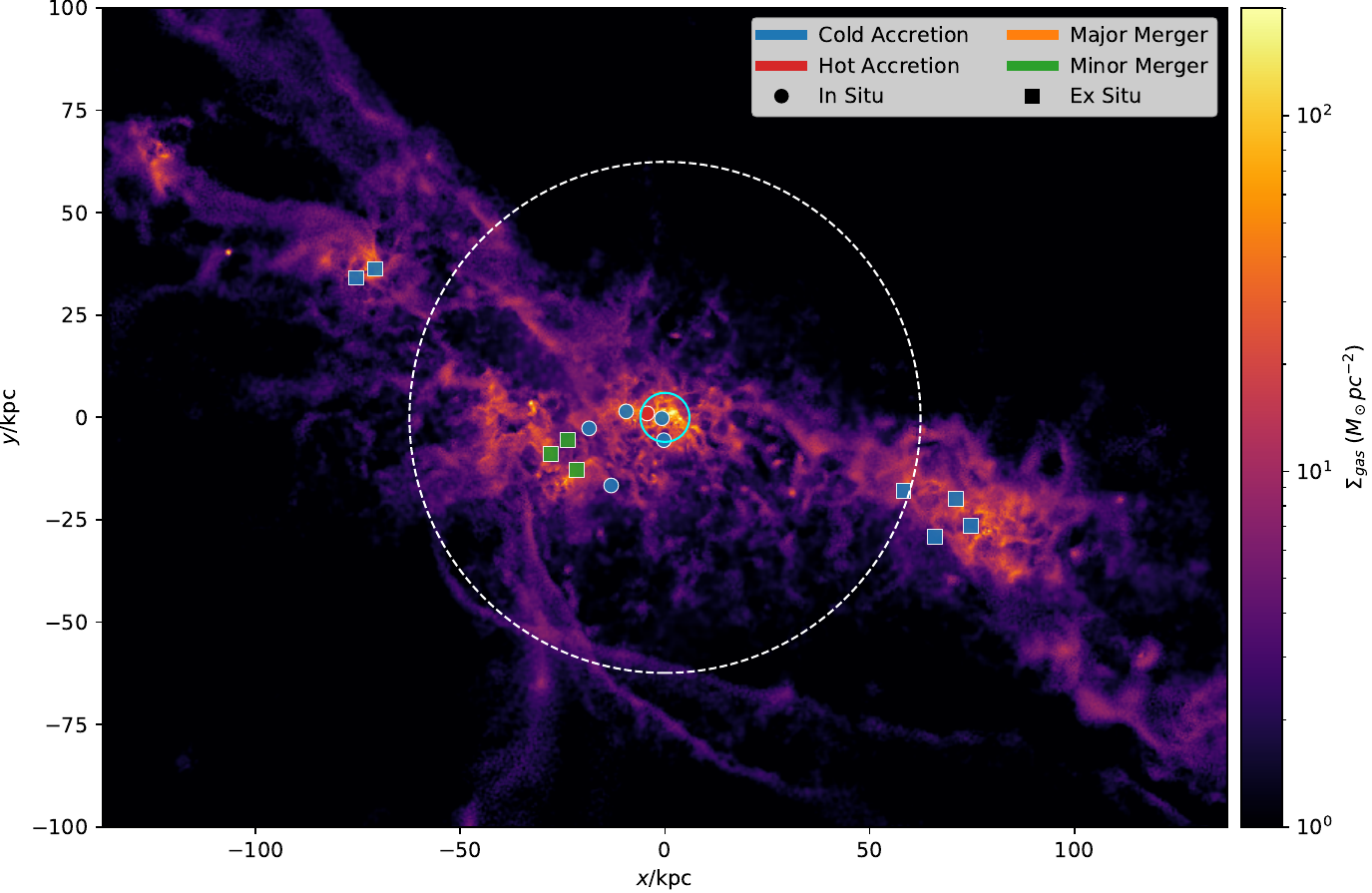}
  \includegraphics[width=0.97\textwidth]{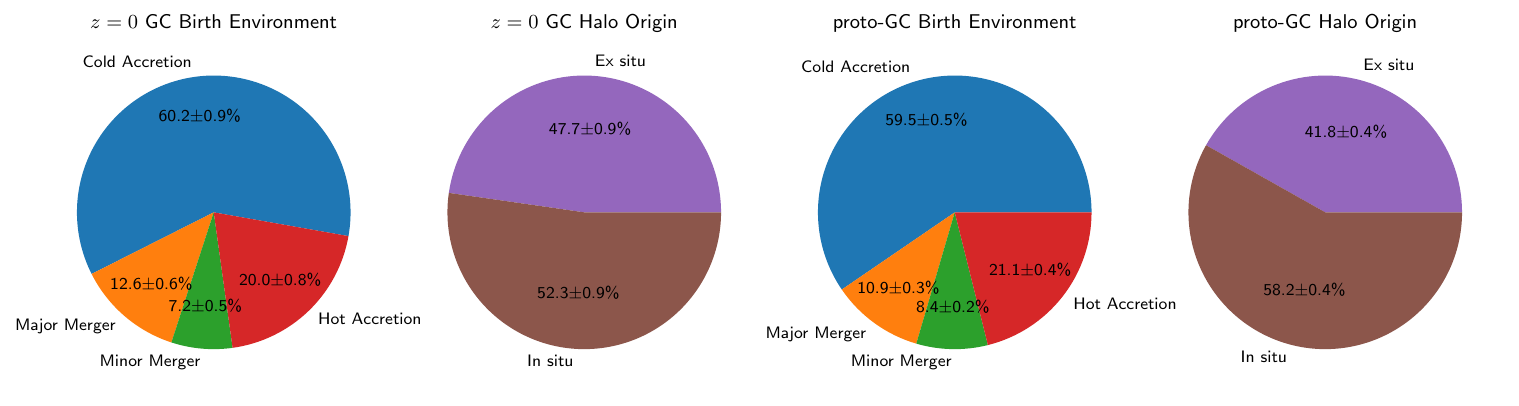}
  \caption{Illustration of the environments where surviving GCs and proto-GCs are predicted to be observed in high-redshift galaxies. The top panel shows a simulation of a typical GC-forming galaxy at redshift $z=2.5$, with colors indicating the gas surface density. Gas particles that are about to form proto-GCs are marked with circles (if they are part of the main progenitor) or squares (if their natal galaxy will merge with the main progenitor later). The particles are colored according to origin of GC formation, with blue indicating gas that was accreted onto the galaxy without ever reaching a temperature above 200,000~K, red indicating gas that was did reach a temperature above 200,000~K, orange indicating formation in a major merger (stellar mass ratio above 1:4), and green indicating formation in a minor merger (stellar mass ratio between 1:4 and 1:10). The dashed white circle indicates the halo's virial radius, and the solid cyan circle indicates the galaxy's stellar half-mass radius. The image illustrates that a minor merger will trigger the formation of three GCs in the disk outskirts, and only a single GC formed from gas that had previously been shock-heated. Six GCs will form in satellite galaxies prior to their accretion onto the main progenitor. The pie charts shown at the bottom of the figure generalize these types of statistics over a suite of 25 cosmological zoom-in simulations of Milky Way-mass galaxies \citep{pfeffer18,kruijssen19d}, for surviving GCs (left) and the full population of proto-GCs (right). For such galaxies, we see that most GCs are predicted to have formed in galaxy disks, with only a minority forming in mergers, and a roughly equal division between formation in the main progenitor and accreted satellites. Minor differences between the statistics of GCs and proto-GCs trace differences in survival rates between the various formation channels. Taken from \citet{keller20}, reproduced with permission.}
  \label{fig:keller20}
\end{figure}
For now, the success at reproducing the observed proto-GC demographics provides some confidence that the physical processes that determine the formation of proto-GCs are broadly consistent with the physical processes that govern star cluster formation in the local Universe, and have likely resulted in the GC populations that we observe at the present day. Based on the predictions of these models, future surveys aiming to study the progenitors of surviving GCs should mostly target the gas-rich disks of galaxies on the star formation main sequence at redshifts $z=1{-}6$, with a high-pressure ISM ($P/k=10^5{-}10^8~\kcmc$) that is fed by cold gas accretion \citep[see \autoref{fig:keller20} and][]{keller20}. While galaxy mergers host considerable YMC formation in the local Universe \citep[e.g.][although most of these might not survive, see \citealt{kruijssen12c}]{schweizer96,whitmore99}, the conditions needed for GC formation are so commonplace at high redshift that mergers are not required, and in fact are even subdominant \citep{keller20}. 

The final empirical piece of the puzzle will arrive with the observations performed by the ELT and Euclid over the coming decade. Respectively, these will deliver GC population demographics at redshifts intermediate to their formation and the present day, and virtually all-sky present-day GC demographics \citep[e.g.][]{marleau24}. Such information is critical for establishing how the initial conditions of the (proto-)GC population and the present-day GC population are related, providing a complete picture of the GC formation, destruction, and evolution from birth to the present day. Indeed, the GC formation framework illustrated in \autoref{fig:architecture} makes a number of strong predictions at intermediate redshifts, mostly related to cluster destruction. Examples are that the characteristic mass scale of the GCMF should be in place at $z\sim1$ due to the early phase of rapid tidal shock-driven cluster destruction \citep[in contrast to earlier models that relied on GC destruction by tidal evaporation, e.g.\ \citealt{fall01}]{kruijssen15b,reinacampos22}, and that the GC system mass-halo mass relation should evolve strongly between redshift $z=0{-}1$ \citep[mostly due to continued galaxy mass growth, but also due to mild evaporation-driven GC destruction, e.g.][]{bastian20}. Additionally, Euclid might help overcome an important sampling bias that exists at the moment, wherein observational studies of present-day GC populations are mostly focused on early-type galaxies and galaxy clusters \citep[e.g.][]{cote04,jordan07b,brodie14,carlsten22,harris23}. However, most galaxies are star-forming \citep[e.g.][]{moustakas13}, most GCs reside in star-forming galaxies \citep{harris16}, and most numerical simulations of GC formation during galaxy formation and evolution model systems that become star-forming main sequence galaxies at $z=0$ \citep[e.g.][]{pfeffer18,grudic23}. Achieving a more representative observational census of GC populations remain an important goal for the near future.

\subsection{Galactic Archaeology: Reconstructing the Assembly Histories of Galaxies with Globular Clusters}
The considerable success of the self-consistent GC formation and destruction framework that has been emerging over the past decade (see \autoref{fig:architecture}), wherein GCs are the natural outcome of high-pressure star formation during galaxy formation and evolution, implies that GC formation models can be used as a baseline for studies in other areas, including e.g.\ gravitational wave astronomy and black hole astrophysics. The most immediate of these applications is galactic archaeology, where GCs can be utilized as fossils tracing the formation and assembly histories of their host galaxies. The quantitative use of GCs for this purpose has always required an end-to-end model for GC formation and evolution, but the recent advances in GC formation theories now satisfy this requirement. If the current GC formation framework holds, then GCs are the relics of the progenitor galaxies that merged to form galaxies like the Milky Way, and certain groups of GCs have come from specific progenitor galaxies.

The use of GCs in galactic archaeology is not new, and has been practiced at least since the seminal paper by \citet{searle78}, who compared the metallicities of GCs in the Milky Way with the fraction of blue stars on their horizontal branches to obtain a proxy for the age-metallicity relation of Galactic GCs. They found that the outer halo contains GCs with a wider range of ages than the inner Milky Way, and suggested that these GCs `originated in transient protogalactic fragments that contined to fall into dynamical equilibrium with the Galaxy for some time after the collapse of its central regions had been completed'. In other words, they used GCs to reveal hierarchical galaxy assembly. With the discovery of the remnant of the Sagittarius dwarf spheroidal galaxy on the other side of the Galactic Center \citep{ibata94}, the association of specific Galactic GCs with its remnant \citep[e.g.][]{bellazzini03}, and their subsequent age measurements \citep{marinfranch09}, it became clear that GCs of accreted dwarf galaxies do indeed have lower metallicities at fixed age than GCs that formed in-situ \citep{forbes10}.

Since then, models of GC formation during galaxy assembly have provided a wide variety of age-metallicity relations \citep[e.g.][]{leaman13,li14,choksi18,kruijssen19d,delucia24}. As expected, these models revealed that the distribution of GCs in age-metallicity space is a strong function of the host galaxy assembly history, with groups of GCs formed in the same progenitor galaxy aligning with the metallicity evolutionary tracks of their accreted natal galaxies. While the age-metallicity information on its own is insufficient to identify common progenitors, this is eased somewhat thanks to exquisite kinematics obtained with the Gaia mission \citep[e.g.][]{gaia18b,vasiliev19,malhan22}. With these data, it has become possible to identify groups of GCs in high-dimensional age-kinematics-chemical composition phase space, revealing which GCs might have formed in the same progenitor galaxy \citep[e.g.][]{massari19,callingham22}.

By using galaxy formation simulations that include GCs to relate the properties of accreted galaxies with the demographics of the GCs they brought along, it has become possible to characterize several of the progenitor galaxies of the Milky Way. This has provided estimates of their masses and accretion redshifts exclusively based on the properties of the Milky Way's GC population \citep{pfeffer20}, culminating in a first reconstruction of the Milky Way's merger tree and showing that the Milky Way assembled unusually rapidly for a galaxy of its mass \citep{kruijssen19e,kruijssen20}. It even led to the identification of a new progenitor galaxy (named `Kraken'), thought to have accreted onto the inner Milky Way some 11~Gyr ago based on its unique signature in age-metallicity space \citep[also see \citealt{massari19} and \citealt{forbes20}]{kruijssen20}. There has been some skepticism regarding the accreted nature of these GCs when using only chemo-dynamic information \citep{belokurov24}, but the inclusion of GC ages makes Kraken's GCs stand out relative to the in-situ component \citep{kruijssen20}. Owing to the combination of Kraken's considerable mass ($\sim3\times10^8~\msun$) implied by the GC ages, metallicities, and orbital binding energies, and the high number of GCs it contributed ($\sim15$), there is space to accommodate only one such progenitor in the Milky Way's central GC population. The stellar debris that has since been identified in the inner Milky Way by \citet{horta21} therefore likely originates from Kraken, which is further supported by its chemo-dynamical similarity to Kraken's GCs. This example is very likely only the beginning -- the continued use of GCs in combination with field star chemodynamics and individual stellar ages \citep[e.g.][]{borre22} is likely to further elucidate the Milky Way's turbulent past. The progress made to date has already contributed to enabling the first numerical simulations of the likely assembly history of the Milky Way \citep{sharpe24}.

Similar successes can plausibly be achieved for external galaxies using age-metallicity-kinematic information, once GC ages are routinely obtained from integrated light \citep[e.g.][]{usher19,cabreraziri22}. More generally, it is now possible to identify which GCs might have formed in-situ and which were accreted from external galaxies to an accuracy of $\sim90\%$ \citep{trujillogomez23}. This means that reconstructing the merger trees of external galaxies using their GC populations is now a realistic possibility, unlocked by the emergence of a comprehensive theoretical framework for the formation and evolution of GCs during galaxy formation.

\subsection{Concluding Perspectives and Avenues for Future Research} \label{sec:perspectives}
Over the past decade or so, a first comprehensive theory for the formation and evolution of GCs within the context of galaxy formation has emerged. In this picture, a population of stellar clusters forms in the cold ISM of high-pressure galaxies, where they are subject to rapid destruction due to tidal shocks from gaseous substructure in the natal galaxy disk. Only the most massive and highest-density clusters manage to survive until they migrate out of the disk (e.g.\ by galaxy merging), thereby enabling their long-term survival and turning them into long-lived GCs. These GCs continue to orbit in the potential well of their host galaxy, where they lose mass and evolve structurally by the comparatively slow effects of two-body relaxation and tidal evaporation. Eventually, this leaves the GC populations that we observe at the present day.

While this picture is supported by the extensive empirical evidence discussed in this review, it is far from complete. The most pressing questions that the author expects will motivate the next decade of research are provided below. Together, these will capitalize on the unique opportunity offered by the next generation of telescopes to for the first time test our understanding of GC formation across the full trinity of GC formation diagnostics (see Section~\ref{sec:trinity}), addressing fundamental physics questions also outside of GC formation (such as star formation, galaxy evolution, cosmology, and black hole research). The questions below all provide an excellent starting point for future studies.
\begin{enumerate}
  \item \textbf{Improve our fundamental understanding of star cluster formation and destruction.} The physical descriptions of these processes have been advancing, but these must reach the point where they become an unambiguous basis on which to build models of GC formation and evolution. The big revolution will be to think in terms of scale-free hierarchies rather than in terms of categorizations. Key questions that we will then be able to address are:
  \begin{enumerate}
    \item What are the typical age spreads of star clusters, and do these vary with the cluster mass?
    \item How can we best exclude gravitationally unbound clusters from extragalactic cluster samples, to obtain a more robust CFE?
    \item How can we accurately infer the low- and high-mass truncation masses of the ICMF, given the observational challenges in terms of completeness and low-number statistics, respectively?
    \item How can we explain the origin of light element abundance variations within GCs in a way that is consistent with our understanding of star formation and stellar feedback, e.g.\ without invoking multiple generations of star formation?
    \item Which physics set the comparatively poorly characterized (initial and final) cluster mass-radius relation?
    \item How can the infinite number of possible tidal histories be standardized to a single framework that enables their systematic study?
    \item How can we empirically quantify the relative contributions of the two main cluster destruction mechanisms (tidal shocks and tidal evaporation) to the mass loss experienced by clusters throughout cosmic history?
  \end{enumerate}
  \item \textbf{Construct a comprehensive theory for the formation and evolution of GCs.} How do the physics of star cluster formation and destruction in the context of galaxy formation and evolution explain the existence of the observed GC population? This is an absolute prerequisite for considering GC formation a solved problem. Key questions that we should address are:
  \begin{enumerate}
    \item What fraction of GCs did \textit{not} form as the product of `normal' cluster formation in high-redshift galaxies?
    \item Are there other formation mechanisms and if so what are they?
    \item What physical processes enable the formation of massive GCs with metallicities below the metallicity floor, i.e.\ $\feh<-2.5$?
    \item If multiple formation mechanisms generated the current GC population, can GC ages measured at high redshift distinguish these?
    \item What was the shape of the initial GC mass function?
    \item How did proto-GCs escape the destructive environment of their host galaxy disk?
    \item What fraction of GCs was destroyed by dynamical friction?
  \end{enumerate}
  \item \textbf{Obtain a complete census of proto-GCs demographics at high redshift, and of the subsequent emergence of the GC population.} This is the step change enabled by the next generation of telescopes, which will allow us to obtain statistically representative samples of (proto-)GCs across the redshift range needed to achieve an end-to-end understanding of GC formation. This will result in a synthesis of the GC population across cosmic time, with potential major implications for star formation, black holes, and gravitational waves. Key questions that we should address are:
  \begin{enumerate}
    \item Given our current understanding of GC formation, can we predict the demographics of the population of (proto-)GCs across the redshift range that will be seen by JWST, Euclid, and 30m class telescopes such as the ELT?
    \item At what redshift did the formation rate of proto-GCs peak?
    \item At what redshift did the formation rate of GCs that eventually survive peak?
    \item Can the unusual chemical properties of multiple populations in GCs generate unusual abundance patterns in high-redshift galaxies, possibly aided by a high CFE or an ICMF with an elevated minimum cluster mass?
    \item How many times more massive were GCs at birth, and how many proto-GCs were there that did not survive to become GCs?
    \item At what redshift were the demographics of the current GC population in place, and how does this depend on the host galaxy mass and assembly history?
    \item Are initial (and final) GC demographics affected by cosmic variance, i.e.\ do we observe the same statistics if we consider fields that fall outside of each others' light cones?
  \end{enumerate}
  \item \textbf{How do GCs trace the assembly histories of their host galaxies?} This major step is unlocked by a comprehensive theory for the formation and evolution of GCs, and our current understanding has enabled the first successful applications of GCs as tracers of galaxy assembly in the Milky Way. The big next step is to industrialize this potential and use GCs to reconstruct the merger trees of external galaxies and perform a fundamental test of cold dark matter cosmology. Key questions that we should address are:
  \begin{enumerate}
    \item How can we further improve the accuracy of GC age measurements from integrated light?
    \item What are the key observables that we need to trace a galaxy's growth and merger history using its GCs?
    \item What are the merger trees of nearby galaxies, as traced by their GCs, as a function of galaxy mass and galaxy clustering?
    \item How do these GC-inferred merger trees compare to the predictions of cold dark matter cosmology?
  \end{enumerate}
\end{enumerate}
This is a highly ambitious set of questions, but it is certainly not beyond the realm of possibility to answer them over the next 10-20 years. The upcoming decade will likely see the first pan-redshift tests of the emerging GC formation theory thanks to revolutionary facilities like JWST, Euclid, and the ELT. These observational studies should focus on measuring the most physically meaningful observables that help distinguish between the many theoretical predictions that now exist. The big challenge for models will remain to match as many observables as possible, and to identify predictions that can meaningfully distinguish between ideas. More than ever, this will require collaboration across the traditional silos of theory and observations. Maybe then we will be able to break the flavor-of-the-decade chronology of GC formation theories, and establish a deep understanding of arguably the most intruiging objects observed across cosmic history.

\begin{ack}[Acknowledgments]{}
  JMDK gratefully acknowledges funding from the European Research Council (ERC) under the European Union's Horizon 2020 research and innovation programme via the ERC Starting Grant MUSTANG (grant agreement number 714907). COOL Research DAO \citep{cool_whitepaper} is a Decentralised Autonomous Organisation supporting research in astrophysics aimed at uncovering our cosmic origins. Ivan Cabrera-Ziri, Nick Choksi, Mike Grudi\'c, Ben Keller, and Joel Pfeffer are acknowledged for kindly permitting the reuse of their figures in this review, and Nate Bastian, M\'{e}lanie Chevance, Chervin Laporte, Steve Longmore, and Joel Pfeffer are thanked for thoughtful feedback on an earlier version of the manuscript. JMDK thanks Nate Bastian, Ivan Cabrera-Ziri, M\'{e}lanie Chevance, Rob Crain, Jim Dale, Gabriella De Lucia, Bruce Elmegreen, Natascha F\"{o}rster Schreiber, Mark Gieles, Oleg Gnedin, Bill Harris, Ben Keller, Mark Krumholz, Henny Lamers, Chervin Laporte, S\o{}ren Larsen, Hui Li, Steve Longmore, Joel Pfeffer, Florent Renaud, Alison Sills, Linda Tacconi, and Simon White for discussions spanning more than 15 years that helped shape his perspective on globular cluster formation and evolution.
\end{ack}

\bibliographystyle{Harvard}
\bibliography{mybib}

\end{document}